\documentclass[format=sigconf,nonacm]{acmart}

\usepackage{relsize}

\usepackage[utf8]{inputenc} 
\usepackage[T1]{fontenc}    
\usepackage{url}            
\usepackage{booktabs}       
\usepackage{amsfonts}       
\usepackage{bm}             
\usepackage{nicefrac}       
\usepackage{microtype}      
\usepackage{xcolor}         

\usepackage{wrapfig}

\usepackage{relsize}

\newif\iftr     
\newif\ifall    
\newif\ifconf   
\newif\ifsq     
\newif\ifnonb   
\newif\iftodos  

\newif\ifsqCAP
\newif\ifsqVS
\newif\ifsqEN
\newif\ifsqTIT
\newif\ifsqR

\newcommand{\tr}[1]{\iftr #1 \fi}

\newcommand{\ignore}[1]{}

\sqtrue
\sqCAPfalse
\sqENtrue
\sqVStrue
\sqTITtrue
\sqRtrue

\sqfalse
\sqCAPfalse
\sqENfalse
\sqVSfalse
\sqTITfalse
\sqRfalse

\usepackage{etex}
\usepackage{balance}
\usepackage{epstopdf}
\usepackage{placeins}

\usepackage{graphicx}
\usepackage{float}
\usepackage{dblfloatfix}
\usepackage{multirow}
\usepackage{rotating}
\usepackage{makecell}
\usepackage{tabulary}
\usepackage{parcolumns}
\usepackage{tikz}
\usetikzlibrary{tikzmark}

\usepackage{xpatch}
\expandafter\xpatchcmd
\csname pgfk@/tikz/every picture/.@cmd\endcsname
{\thepage}{\arabic{page}}{}{}

\tikzstyle{comment} = [draw, fill=blue!70, text=white, text width=3cm, minimum height=1cm, rounded corners, align=left, font=\scriptsize]
\tikzstyle{background_alg} = [draw, fill=blue!20, opacity=0.4, inner sep=4pt, rounded corners=2pt]

\usetikzlibrary{shapes}
\usetikzlibrary{plotmarks}
\usetikzlibrary{calc, fit}

\usepackage{enumitem}

\usepackage{amsthm}
\usepackage{amsmath,amsfonts}
\usepackage{mathtools,mathrsfs}

\usepackage{soul}
\usepackage{fontawesome}
\usepackage{pifont}
\usepackage{textcomp}
\usepackage{booktabs}
\usepackage{url}
\usepackage{pbox}
\usepackage[normalem]{ulem}
\usepackage[10pt]{moresize}

\newcommand{\macb}[1]{\textbf{{#1}}}

\ifsqCAP
\usepackage[font={normalfont, footnotesize}]{caption}
\usepackage[font={normalfont, footnotesize}]{subcaption}
\else
\usepackage[font={normalfont, footnotesize}]{caption}
\usepackage[font={normalfont, footnotesize}]{subcaption}
\fi

\newcommand{\vspaceSQ}[1]{\ifsqVS\vspace{#1}\fi}
\newcommand{\enlargeSQ}[1]{\ifsqEN\enlargethispage{\baselineskip}\fi}

\ifsqTIT
\usepackage[compact]{titlesec}
\titlespacing*{\section}{2pt}{3pt}{2pt}
\titlespacing*{\subsection}{2pt}{2pt}{1pt}
\titlespacing*{\subsubsection}{2pt}{3pt}{1pt}
\fi

\usepackage{xcolor}

\definecolor{brick}{RGB}{205,92,92}

\definecolor{lgrey}{RGB}{230,230,230}
\definecolor{lred}{RGB}{255,200,200}
\definecolor{lblue}{RGB}{200,200,255}
\definecolor{lyellow}{RGB}{255,255,100}
\definecolor{lgreen}{RGB}{144,238,144}

\definecolor{llgrey}{RGB}{240,240,240}
\definecolor{llred}{RGB}{255,220,220}
\definecolor{llblue}{RGB}{230,230,255}
\definecolor{llyellow}{RGB}{255,255,220}
\definecolor{llgreen}{RGB}{230,255,230}

\definecolor{darkgrey}{RGB}{70,70,70}
\definecolor{lightgrey}{RGB}{200,200,200}
\definecolor{lyellow}{RGB}{255,255,100}
\definecolor{llyellow}{RGB}{250,250,180}
\definecolor{lgreen}{RGB}{144,238,144}
\definecolor{raphael_comments}{RGB}{13, 145, 24}

\usepackage[customcolors]{hf-tikz}
\hfsetbordercolor{white}
\hfsetfillcolor{vlgray}

\definecolor{vlgray}{rgb}{0.77 0.77 0.77}
\definecolor{ablack}{rgb}{0.2 0.2 0.2}
\definecolor{vllgray}{rgb}{0.9 0.9 0.9}
\definecolor{bblue}{rgb}{0.7 0.7 0.99}

\definecolor{darkgreen}{RGB}{0,127,0}
\definecolor{darkred}{RGB}{127,0,0}
\definecolor{shadecolor}{RGB}{250, 250, 250}

\usepackage{colortbl}

\usepackage{listings}

\ifsq
\else
\fi

\lstdefinestyle{python}{ %
	language=Python,
	frame=tb, 
	framerule=0pt,
	basicstyle=\tt\ssmall, 
	backgroundcolor=\color{shadecolor},
	keywordstyle=\color{blue}\bfseries,
	commentstyle=\color{darkgreen},
	rulecolor=\color{black},
	stringstyle=\color{darkred},
	lineskip=0pt,
	keywords={def, @dace, program, map, lambda, import, numpy, np, int32, float32, float64, complex128, with, as, True, False, return, for, in, dc, nb, numba, dace, cupy, cp, @numba, @nb, @dc, symbol, mpi, deinsum},
	numbers=none,
	numbersep=5pt,
	showstringspaces=false
}

\newcommand{\maciej}[1]{\textcolor{blue}{[Maciej: #1]}}

\definecolor{hlL}{rgb}{0.8 0.8 0.99}

\newcounter{highlight}

\newcounter{hlLR}

\newcounter{hlLIR}

\newcounter{hlLIIR}

\newcounter{Ahighlight}

\renewcommand{\epsilon}{\ensuremath\varepsilon}

\renewcommand{\phi}{\ensuremath{\varphi}}

\usepackage{stmaryrd}

\usepackage{yfonts}

\usepackage{mathalfa}

\newcommand{\faY}[0]{\faBatteryFull}
\newcommand{\faH}[0]{\faBatteryHalf}
\newcommand{\faN}[0]{\faTimes}

\renewcommand{\arraystretch}{1.0}

\trtrue
\conffalse

\usepackage{colortbl}

\newcommand{\marginparN}[1]{\marginpar{}}
\newcommand{\marginparX}[1]{}

\usepackage{float}

\usepackage[skins]{tcolorbox}
\tcbset{commonstyle/.style={boxrule=0pt,sharp corners,enhanced jigsaw,nobeforeafter,boxsep=0pt,left=\fboxsep,right=\fboxsep}}
\newtcolorbox{mycolorbox}[1][]{commonstyle,#1}

\usepackage{amsbsy}

\iftr

\renewcommand\footnotetextcopyrightpermission[1]{} 
\setcopyright{none}

\fi

\begin{document}

\title{The Graph Database Interface: Scaling Online Transactional and Analytical Graph Workloads to Hundreds of Thousands of Cores}

\author{Maciej Besta$^{1*\#}$,
Robert Gerstenberger$^{1*\#}$,
Marc Fischer$^1$,
Michał Podstawski$^{2,3}$,\\
Nils Blach$^1$,
Berke Egeli$^1$,
Georgy Mitenkov$^1$,
Wojciech Chlapek$^4$,\\
Marek Michalewicz$^5$,
Hubert Niewiadomski$^6$,
J\"{u}rgen M\"{u}ller$^7$,
Torsten Hoefler$^1$$^*$}
       \affiliation{\vspace{0.3em}$^1$ETH Zurich;
       {$^2$}TCL Eagle Lab;
       {$^3$}Warsaw University of Technology;
       {$^4$}ICM UW;\\
       {$^5$}Sano Centre for Computational Medicine;
       {$^6$}Cledar;
       {$^7$}BASF SE;
       {$^*$}Corresponding authors, 
       {$^\#$}alphabetical order\\
}

\renewcommand{\shortauthors}{M. Besta et al.}

\renewcommand\abstractname{ABSTRACT}

\begin{abstract}
Graph databases (GDBs) are crucial in academic and industry applications.  The
key challenges in developing GDBs are achieving high performance, scalability,
programmability, and portability. To tackle these challenges, we harness
established practices from the HPC landscape to build a system that outperforms
all past GDBs presented in the literature by orders of magnitude, for both OLTP
and OLAP workloads. For this, we first identify and crystallize
performance-critical building blocks in the GDB design, and abstract them into
a portable and programmable API specification, called the Graph Database
Interface (GDI), inspired by the best practices of MPI. We then use GDI to
design a GDB for distributed-memory RDMA architectures. Our implementation
harnesses one-sided RDMA communication and collective operations, and it offers
architecture-independent theoretical performance guarantees. The resulting
design achieves extreme scales of more than a hundred thousand cores. Our work
will facilitate the development of next-generation extreme-scale graph
databases.
%
%
%
\end{abstract}

\settopmatter{printfolios=false,printccs=false,printacmref=false}

\maketitle

{\normalsize\noindent\macb{Code and GDI Specification:} {\small\url{https://github.com/spcl/GDI-RMA}}}

\section{INTRODUCTION}
\label{sec:intro}


\begin{table*}[t]
\vspaceSQ{-1em}
\centering
\ifsq
\setlength{\tabcolsep}{1.5pt}
\else
\setlength{\tabcolsep}{1.5pt}
\fi
\footnotesize
\ifsq\renewcommand{\arraystretch}{0.7}\fi
\ifsq
\scriptsize
\else
\footnotesize
\fi
\begin{tabular}{lllllllllrrrrrrrrrrrrl@{}}
\toprule
\multirow{2}{*}{\textbf{Reference}} & \multirow{2}{*}{\textbf{RDMA?}} & \multirow{2}{*}{\textbf{Prog.?}} & \multicolumn{2}{c}{{\textbf{Port.?}}} & \multicolumn{4}{c}{\textbf{Focus on...}} & \multicolumn{5}{c}{\textbf{Achieved scales (OLTP)}} & \multicolumn{5}{c}{\textbf{Achieved scales (OLAP, OLSP)}}  & \multirow{2}{*}{\textbf{MemS?}} & \multirow{2}{*}{\textbf{Th.?}} \\
\cmidrule{4-5} \cmidrule{6-9} \cmidrule{10-14} \cmidrule{15-19}
& & & \textbf{wR} & \textbf{bR} & \textbf{OLTP} & \textbf{OLAP} & \textbf{OLSP} & \textbf{BULK} & \textbf{\#S} & \textbf{\#C} & \textbf{Size} & \textbf{$|E|$} & \textbf{$|V|$} & \textbf{\#S} & \textbf{\#C} & \textbf{Size} & \textbf{$|E|$} & \textbf{$|V|$} & & \\
\midrule
A1~\cite{buragohain2020a1}  & \faY & \faN & \faN &\faN & \faY & \faN & \faN & \faN & 245 & 2,940 & 3.2 TB & 6.2B & 3.7B & \faN & \faN & \faN & \faN & \faN & 128 GB & \faN \\
GAIA~\cite{qian2021gaia}  & \faN & \faN & \faN & \faN & \faN & \faY & \faN & \faN & \faN & \faN & \faN & \faN & \faN & 16 & 384 & 1.96 TB & 17.79B & 2.69B & 512 GB & \faN \\
G-Tran~\cite{chen2021g} & \faY & \faN & \faN & \faH & \faY & \faN & \faN & \faN & 10 & 160 & $^*$1.28 TB & 0.495B & 0.082B & \faN & \faN & \faN & \faN & \faN & 128 GB & \faH \\
Neo4j~\cite{janez2020basf} & \faN & \faN & \faN & \faN & \faY & \faY & \faH & \faY & 1 & 128 & 6.9 TB & 55B & 5B & 1 & 128 & 6.9 TB & 55B & 5B & 2 TB & \faN \\
TigerGraph~\cite{tigergraph2022ldbc} & \faN & \faN & \faN & \faN & \faY & \faY & \faN & \faY & 40 & 1600 & 17.7 TB & 533.5B & 72.62B & 36 & 4,608 & N/A & 539.6B & 72.6B & 1 TB & \faN \\
JanusGraph~\cite{janus_graph_links} & \faN & \faN & \faN & \faN & \faY & \faH & \faN & \faY & N/A & N/A & N/A & N/A & N/A & N/A & N/A & N/A & N/A & N/A & N/A & \faN \\
Weaver~\cite{dubey2016weaver} & \faN & \faN & \faN & \faN & \faY & \faN & \faN & \faN & 44 & 352 & 0.976 TB & 1.2B & 0.08B & \faN & \faN & \faN & \faN & \faN & 16 GB & \faN \\
Wukong~\cite{shi2016wukong} & \faY & \faN & \faN & \faN & \faY & \faN & \faN & \faN & 6 & 120 & $^*$0.384 TB & 1.41B & 0.387B & \faN & \faN & \faN & \faN & \faN & 64 GB & \faN \\
ByteGraph~\cite{libytegraph} & \faN & \faN & \faN & \faN & \faY & \faY & \faY & \faN & 10 & 160 & N/A & N/A & N/A & 130 & N/A & 113 TB & N/A & N/A & 1 TB & \faH \\
\midrule
%
\textbf{This work} & \faY & \faY & \faY & \faY & \faY & \faY & \faY & \faY & \textbf{7,142} & \textbf{121,680} & \textbf{77.3 TB} & \textbf{549.8B} & \textbf{34.36B} & \textbf{7,142} & \textbf{121,680} & \textbf{36.5 TB} & \textbf{274.9B} & \textbf{17.2B} & \textbf{64 GB} & \faY \\
\bottomrule
\end{tabular}
\caption{\textmd{Comparison of graph databases.
``\textbf{RDMA?}'': Is a system primarily targeting RDMA architectures?
``\textbf{Prog.?}'': Does a system's storage and transactional backend design focus on programmability and code simplicity?
``\textbf{Port.?}'': Does a system foster portability? If yes, is it portability
within different RDMA architectures (``\textbf{wR}''), or also beyond RDMA (``\textbf{bR}'')? 
``\textbf{Supported workloads}'': What are supported workloads? (all workloads are explained in Section~\ref{sec:workloads}) 
``\textbf{Achieved scales}'': What are achieved scales?
``\textbf{\#S}'': Number of servers. 
``\textbf{\#C}'': Number of cores.
``\textbf{Size}'': Total size (in memory) of the processed graph.
``\textbf{$|E|$}'': Number of edges. 
``\textbf{$|V|$}'': Number of vertices.
``\textbf{MemS}'': Amount of memory available in a single server.
``\textbf{Th.?}'': Does a system come with theoretical analysis and
support for its performance and scalability properties?} 
``\faY'': full support,
``\faH'': partial support,
``\faN'': no support,
``$^*$'': \textmd{Estimate.}
\textmd{\emph{The GDI-based system is the only one
that focuses on all major aspects of the GDB design: programmability,
portability, high performance, and very large scales}}.}
\vspace{-2.5em}
\label{tab:scale}
\vspace{-0.5em}
\end{table*}

Graph databases (GDBs) enable storing,
processing, and analyzing large and evolving irregular graph
datasets in areas as different as medicine or sociology~\cite{besta2019demystifying, miller2013graph}. 
GDBs face unique design and compute challenges.
First, GDB datasets are huge \emph{and} complex. 
While they can have over tens of
  trillions of edges~\cite{lin2018shentu}, both vertices and edges may also come with
  arbitrarily many labels and properties. This further increases
  dataset sizes. On top of that,
  \emph{much} larger datasets are already on the horizon\footnote{As indicated
  by discussions with our industry partners}.
Second, while traditional GDB workloads focus on online transactional
processing (OLTP), there is a growing interest in supporting other classes such
as online analytical processing (OLAP) or the ``online serving processing''
(OLSP), also called business intelligence~\cite{libytegraph}.
\emph{How to design {high-performance and
scalable} databases that enable processing of {large and complex} graphs for
{OLTP, OLAP, and OLSP} queries?}
Third, portability is also important - there are many different hardware
architectures available, and it may be very tedious and expensive to port a
database codebase to each new class of hardware. 
%
%
Finally, a GDB design that would satisfy all the above challenges may become
extremely complicated, and consequently hard to reason about, debug, maintain,
or extend. This raises the question: \emph{How to ensure {portability} and
{programmability} of complex next-generation graph databases, without
compromising on their performance?}

To resolve \emph{all} the above challenges, we provide \emph{the first
principled approach for designing and implementing large-scale GDBs}. This approach harnesses some
of the most powerful practices and schemes from the HPC domain,
several of them for the first time in GDB system design.
Our approach is inspired by the Message-Passing Interface (MPI)~\cite{mpi3} and
numerous successes it has in designing and developing portable, programmable,
high-performance, and scalable applications.
We propose to approach the GDB design in a similar way: (1) identify
performance-critical building blocks, (2) build a portable API, (3) implement
this API with high-performance techniques such as collectives or one-sided
RDMA, and (4) use the API implementation to build the desired GDB system.
In this work, we execute these four steps, and as a result we deliver a
publicly-available GDB system that resolves all the four challenges.

First, we analyze the design and codebases of many
GDBs~\cite{besta2019demystifying} (e.g., Neo4j~\cite{neo4j_book}, Apache
TinkerPop~\cite{tinkergraph_links}, or JanusGraph~\cite{janus_graph_links}) to
identify fundamental performance-critical building blocks. We then crystallize
these blocks into a portable and programmable specification called the Graph
Database Interface (GDI) 
(\textbf{contribution~\#1}).
%
%
GDI focuses on the data storage layer, covering database
transactions, indexes, graph data, graph metadata, and others.
GDI is portable because -- as MPI -- \emph{it is fully decoupled from its implementation}.
\iftr
Besides providing some advice to implementors, GDI does not specify any
architecture- or implementation-specific details, instead focusing on a clear
specification of the semantics of its routines.
\fi
Hence -- just like with MPI-based applications -- any database based on GDI
could be seamlessly compiled and executed on any system, if there is
a GDI implementation for that system.

\enlargeSQ
\enlargeSQ

Second, we offer a high-performance
implementation of GDI for distributed-memory (DM) systems supporting RDMA-enabled
interconnects, called GDI-RMA.
We use GDI-RMA to build a
highly-scalable GDB engine (\textbf{contribution~\#2}).
\iftr
We focus on DM systems as they offer large amounts of total memory (even up to
10PB~\cite{cao2022scaling}), compute power (even more than 40 million
cores~\cite{cao2022scaling}), and network bandwidth (even 1.66 PB/s and 46.1
TB/s for the injection and bisection bandwidth,
respectively~\cite{ajima2018tofu}). Hence, they can accommodate the enormous
requirements posed by GDB datasets, keeping data fully in-memory to avoid
expensive disk accesses.
\fi
\iftr
Simultaneously, RDMA has been the enabler of scalability and high performance
in both the supercomputing landscape and -- more recently -- in the cloud data
center domain~\cite{recio2007remote, fompi-paper,
mitchell2013using, wang2015hydradb, kalia2014using, simpson2020securing,
huang2012high, islam2012high, lu2013high, woodall2006high, poke2015dare,
liu2004high, kalia2016design, besta2015active,
besta2014fault, schmid2016high}.
\fi
\iftr
RDMA is widely supported on DM systems, both in the cloud
infrastructure (e.g., on Azure~\cite{karmarkar2015availability},
Alibaba~\cite{alibaba_cloud}, or Oracle~\cite{oracle_cloud} data centers), in
general Ethernet networks (with RoCE~\cite{roce} or iWARP~\cite{iwarp}), and in
the HPC networks (e.g., on Cray~\cite{aries}, IBM~\cite{arimilli2010percs},
or InfiniBand~\cite{IBAspec}). Importantly, the RDMA API
in most of these systems can be abstracted into a generic \emph{Remote Memory
Access (RMA)} programming model~\cite{fompi-paper}.
Our implementation of GDI, called \emph{GDI for Remote Memory Access
(GDI-RMA)}, facilitates portability on such systems as it uses this generic RMA
API.
\fi

In GDI-RMA, we make three underlying design decisions for highest performance
and scalability. First, we \emph{carefully design a scalable distributed
storage layer called blocked graph data layout (BGDL)} to enable a tradeoff
between the needed communication and storage. Second, we incorporate the highly
scalable \emph{one-sided non-blocking RDMA communication (puts, gets, and
atomics)}. Third, we use \emph{collective communication
(collectives)}~\cite{chan2007collective, hoefler2014energy}
to deliver scalable transactions involving many processes (e.g.,
for large-scale OLAP queries) with well-defined semantics.
\iftr
Collectives are a form of high-performance data exchange that has been tuned
over many years, resulting in schemes that are provably optimal or near-optimal
(in minimizing the communicated data) and that enable efficient use of all
compute resources~\cite{thakur2005optimization, snir1998mpi,
hoefler2014energy}. 
\fi

We support nearly any function in our implementation with a theoretical
performance analysis that is independent of the underlying hardware
(\textbf{contribution \#3}). This facilitates the reasoning about the
performance and scalability of our GDI implementation.

We illustrate how to use GDI to program many graph database workloads
(\textbf{contribution \#4}), covering OLTP, OLAP, and OLSP.
We consider recommendations by the LDBC and LinkBench academic and industry
benchmarks~\cite{ldbc, armstrong2013linkbench}. We use established
problems such as BFS~\cite{page1999pagerank,
besta2017push} and state-of-the-art workloads such as Graph Neural
Networks~\cite{wu2020comprehensive, zhou2020graph, scarselli2008graph,
zhang2020deep, hamilton2017representation, bronstein2017geometric,
kipf2016semi, gianinazzi2021learning, besta2022parallel}.
Moreover, as there are no publicly available graph datasets with labels and
properties of that magnitude, we also develop an in-memory distributed
generator that can rapidly create such a graph of arbitrary size and
configuration of labels and properties (\textbf{contribution~\#5}).

The evaluation of GDI-RMA (\textbf{contribution~\#6}) significantly surpasses
in scale previous GDB analyses in the literature in the counts of servers,
counts of cores, and in the size of a single analytic workload (see
Table~\ref{tab:scale}).  
We successfully scale to \emph{121,680 cores (7,142 servers)}, using all the 
available memory, and the only reason why we did not try more is because we do
not have access to a larger system. Based on our analysis, we expect that our
GDI implementation could easily achieve the scale of hundreds of thousands of
cores.
We also achieve high throughput and low latencies, 
outperforming modern graph databases such as Neo4j or JanusGraph by more than an order of magnitude in
both metrics.
Our implementation is publicly available (\textbf{contribution~\#7}) to help
achieve new frontiers for GDBs running on petascale and exascale data centers
and supercomputers.

We compare our work to other GDBs in Table~\ref{tab:scale}.
GDI is the only RDMA-based system to support all three fundamental
workloads (OLTP, OLAP, OLSP), and the only one to focus on portability \&
programmability, and with theoretical performance guarantees.

\section{GRAPH DATA MODEL \& WORKLOADS}
\label{sec:back}

We first present basic concepts and notation.
%

\paragraph{Graph Data Model}
\iftr
We target graphs modeled with the established \textbf{Labeled
Property Graph Model (LPG)}~\cite{besta2019demystifying} (also called the
property graph~\cite{gdb_query_language_Angles}). LPG is a primary data model
used in many GDBs, including the leading Neo4j GDB~\cite{besta2019demystifying}. 
\fi
An LPG graph can formally be modeled as a tuple $(V,E,L,l,K,W,p)$.
$V$ is a set of vertices and $E \subseteq V \times V$ is a set of edges;
$|V|=n$ and $|E|=m$. 
\iftr
An edge $e=(u,v)\in E$ is a tuple of two vertices, where $u$ is the out-vertex
(origin) and $v$ is the in-vertex (target). If $G$ is undirected, an edge
$e=\{u,v\}\in E$ is a set of two vertices.
\fi
%
%
%
%
$L$ denotes the set of \emph{labels} that differentiate subsets
of vertices and edges. $l$ is a labeling function, which maps vertices
and edges to subsets of labels; $l : V \cup E \mapsto \mathcal{P}(L)$ with
$\mathcal{P}(L)$ being the power set of $L$, meaning all possible subsets of
$L$. 
Each vertex and edge can also feature
arbitrarily many \emph{properties} (sometimes referred to as attributes). A
property is a $(key, value)$ pair, where the $key$ works as an identifier with
$value$ is the corresponding value.
$\mathnormal{K}$ and $\mathnormal{W}$ are sets of all possible keys and
values, respectively. $p : (V \cup E) \times \mathnormal{K} \mapsto
\mathnormal{W}$ maps each vertex and edge to their properties, given the key. 
We also refer to the elements of $K$ as \emph{property types} (\emph{p-types}).
Note that we distinguish between {property types} and \emph{properties}, the
latter being the specific key-value property tuples attached to vertices/edges.
%
%
%
%

\paragraph{Graph Data vs.~Graph Metadata}
We collectively denote labels~$L$, property types~$K$ and property values~$W$ as the
\textbf{graph metadata} because these sets do not describe any
specific graph elements, but they define the potential labels, keys, and
values. 
A collective name \textbf{graph data} refers to the actual graph
elements, described by $V, E, l, p$.

\paragraph{Graph Database Workloads}
The established LDBC and LinkBench academic and industry
benchmarks~\cite{ldbc, armstrong2013linkbench} identify two main classes of
graph database workloads targeting the LPG graph model: \emph{interactive
workloads}~\cite{ldbc_snb_specification} (mostly OLTP) and \emph{graph
analytics}~\cite{ldbc_graphanalytics_paper} (mostly OLAP).
Interactive workloads are further divided into \emph{short read-only queries}
(which often start with a single graph element such as a vertex and lookup its
neighbors or conduct small traversals) and \emph{transactional updates} (which
conduct simple graph updates, such as inserting an edge).
Next, preliminary efforts also distinguish an additional class of
\emph{business intelligence workloads} (BI)~\cite{early_ldbc_paper,
DBLP:journals/pvldb/SzarnyasWSSBWZB22} which fetch large parts of a graph and
often use data summarization and aggregation.  They are sometimes referred to
as \emph{Online Serving Processing} (OLSP)~\cite{libytegraph}.
Finally, we also distinguish workloads associated with bulk data ingestion
(BULK).  They take place, e.g., when inserting new batches of data into the
system.

\section{THE GRAPH DATABASE INTERFACE} 

\enlargeSQ

GDI is a storage layer interface for GDBs, offering CRUD (create,
read, update, delete) functionality for the elements of the LPG model:
vertices, edges, labels, and properties. The interface provides rich semantics
and transaction handling. The focus of GDI lies on enabling {high-performance},
scalable, and portable implementations of the provided methods.  
\iftr
Thus, higher-level parts of a graph database (query methods, query planer,
execution engine, etc.) can run vendor agnostic. 
\fi
Moreover, GDI facilitates {programmability} by offering a structured set of
routines with well-defined semantics.
%

In this paper, we provide a comprehensive summary of the most important aspects
of GDI. 
We also distill the key design choices and insights beyond the common system
knowledge, that we indicate with the ``\includegraphics[scale=0.2,trim=0 16 0 0]{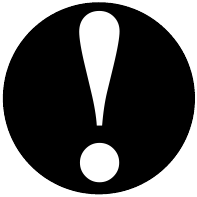}''
symbol.
The full \textbf{GDI specification} is available in a separate
manuscript~\cite{besta2023gdi}.
It contains a detailed description of routines, extensive advice for users and
implementors, naming conventions, description of basic datatypes, and others.

\subsection{Relation Between GDI and Graph Databases}

We illustrate the relation between GDI and a generic graph database landscape
in Figure~\ref{fig:overview}. 
GDI is to be used primarily by the database middle layer, as a storage and
transactional engine. 
%
%
Here, the \textbf{client} first queries the GDB using a graph query language
such as Cypher~\cite{green2019updating}.
Second, the \textbf{database mid-layer} coordinates the execution of the client
query. This could include distributing the workload among multiple machines, or
aggregating as well as filtering intermediate results that ran on different
processes.
The mid-layer relies on the underlying \textbf{storage and transaction engine,
where GDI resides}. This part accesses the graph data and translates from
generic graph-related objects needed by queries to hardware dependent storage.
Therefore, the layer provides a rich set of interfaces to create, read, update,
and delete (CRUD) vertices, edges, labels, and properties, and to
execute transactions. 
Finally, the \textbf{storage backend} provides access to the actual
storage such as distributed RAM, using formats such as CSV files or JSON.
\iftr
Its goal is to store the data in a reliable way and provide fast data access. 
\fi

\iftr
\begin{figure}[b]
\centering
\vspaceSQ{-2.3em}
\includegraphics[width=0.89\columnwidth]{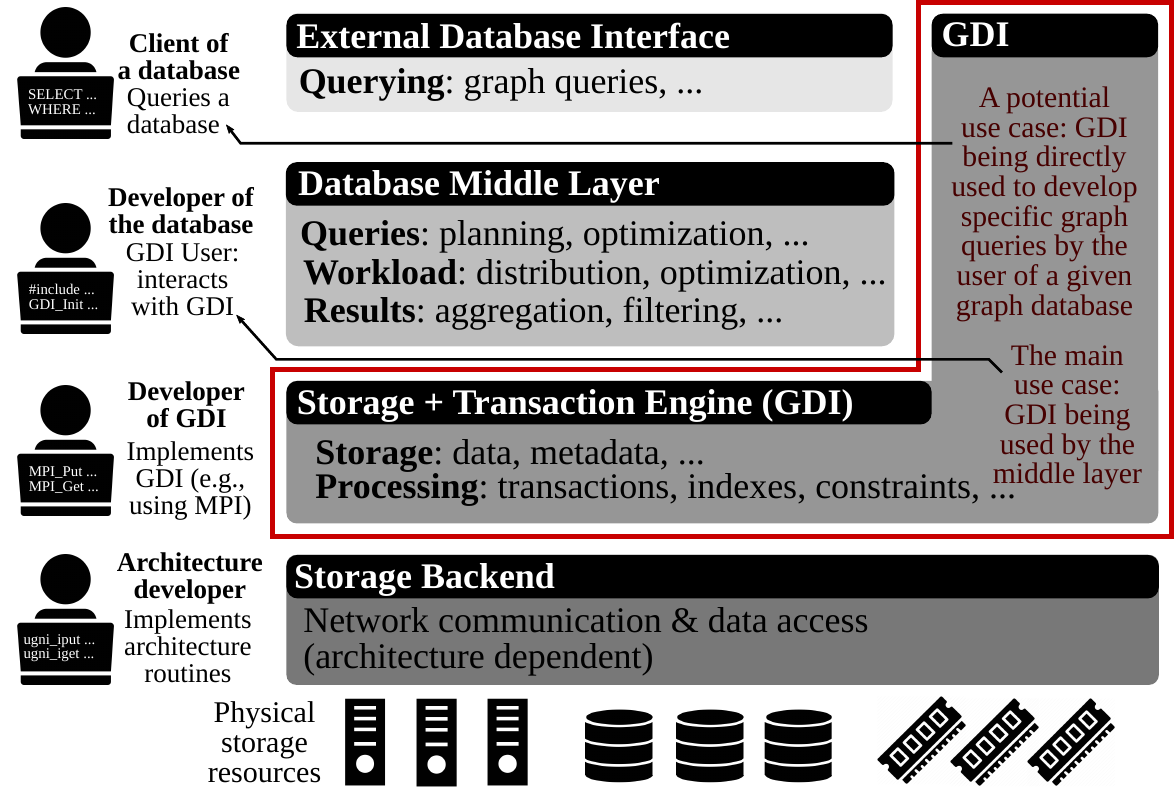}
\vspaceSQ{-1em}
\caption{GDI with respect to other parts of the graph database landscape.}
\vspaceSQ{-1em}
\label{fig:overview}
\end{figure}

\fi

\begin{figure*}[t]
\centering
\ifsq
\includegraphics[width=0.9\textwidth]{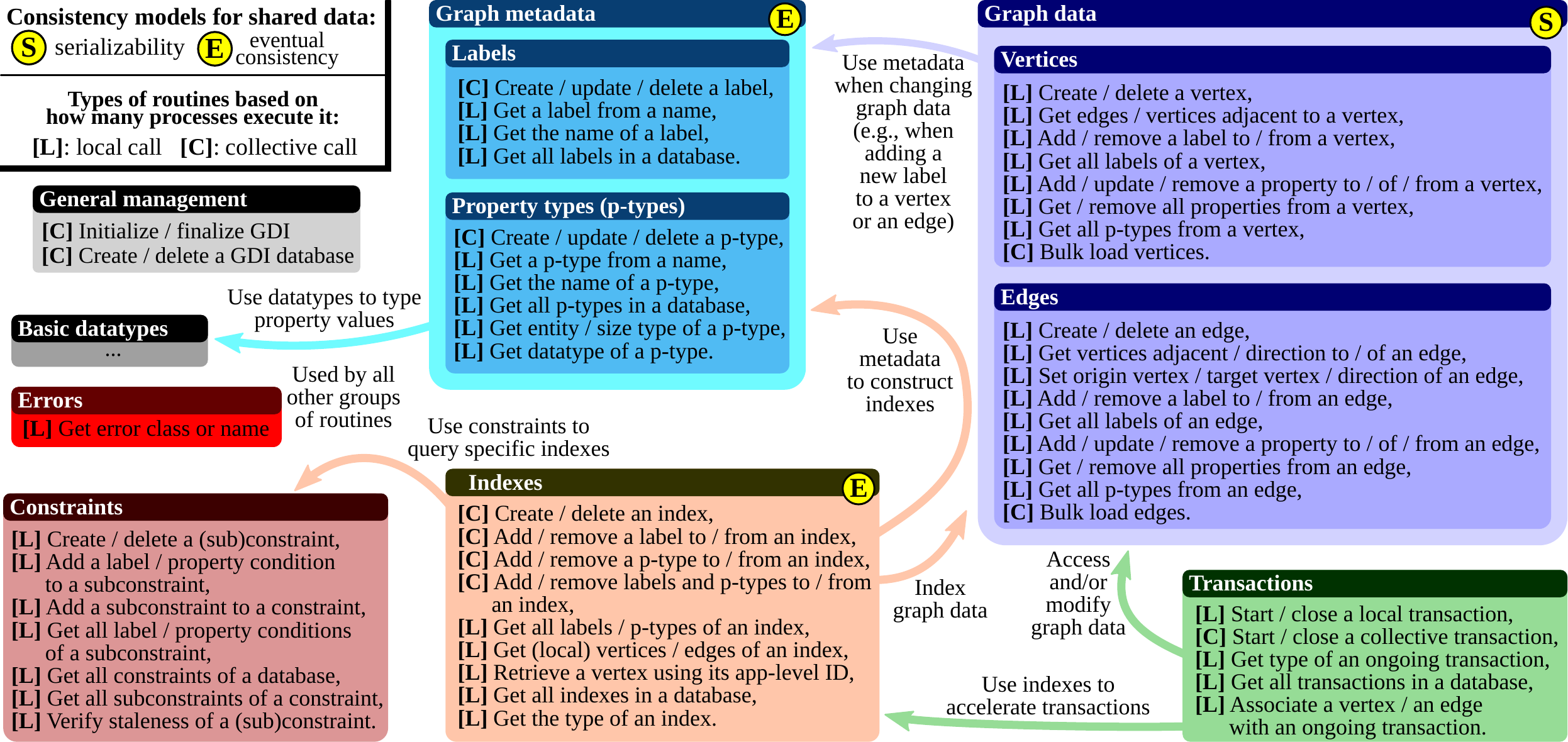}
\else
\includegraphics[width=1.0\textwidth]{gdi-dets_details.pdf}
\fi
\vspaceSQ{-1em}
\caption{Illustration of the classes of GDI routines.}
\vspaceSQ{-1em}
\label{fig:gdi-structure}
\end{figure*}

We also envision that GDI could be used directly by a client, to
directly implement a given query. For this, we will illustrate how to implement
different GDB workloads with GDI in Section~\ref{sec:workloads}.

%
\iftr
\label{sec:cypher_query}
As an example, consider the following query: ``\emph{How many people are over 30
years old and drive a red car?}''. A corresponding Cypher query
by the user could be as follows: \texttt{MATCH (per:Person) WHERE per.age $>$ 30
AND per-[:OWN]-$>$ vehicle(:Car) AND vehicle.color = red RETURN count(per)}.
The mid-layer decides on the necessary steps, which would be retrieving
``Person'' vertices (possibly using an index), checking the age condition,
retrieving edges with the ``OWN'' label, checking the adjacent vertices for the
``Car'' label, verifying if the found ``Car'' vertices are red.
\tr{If an index is available that is not only associated with the label 'Car'
but also the property type 'color', the query planer might instead decide to
start at the other side of the data flow and follow-up in the opposite
direction. That would be a logical step under the assumption, that there are
fewer red-painted cars than persons in the database, so the amount of initial
objects would be (much) smaller. However in the following paragraphs
illustrating the different layers, the first scenario is assumed.}
Next, the mid-layer runs these steps, coordinating each process to examine
their local portions of ``Person'' vertices and to check whether they fulfill
all of the conditions set out above.  The sum of the vertices can be
accumulated locally and the global result is computed with a global reduce
operation.
To execute the above, the mid-layer performs a local index query to retrieve
all the vertices with the label ``Person'', and translate the condition steps
of the query planner into a series of function calls to the storage engine. For
example, to check the age condition, one would first retrieve the ``age''
property of the respective vertex, and afterwards check whether that property
fulfills the age condition. For the edge retrieval, it is possible to let the
storage handle the filtering with the use of a specifically crafted so-called
constraint object.
The storage engine layer receives the function calls from the mid-layer, and it
handles the retrieval of data from the intermediate data representation of the
(possibly distributed) storage backend layer.
Finally, the storage backend directly accesses physical storage as instructed
by the storage engine. 
\fi

\subsection{Structure and Functionalities of GDI}

The GDI interface is structured into groups of routines,  detailed in
Figure~\ref{fig:gdi-structure}.
General GDI and database management schemes (\textcolor{gray}{\textbf{gray
color}}) perform setup needed for any other GDI functions to be able to run.
Graph metadata routines (\textcolor{cyan}{\textbf{cyan color}}) enable
creating, updating, deleting, and querying different aspects of labels and
property types.
Graph data routines (\textcolor{blue}{\textbf{blue color}}) provide CRUD
capabilities for vertices and edges, also including adding, removing, updating,
and querying labels and properties of specific vertices and edges, and bulk
data loading.
Transaction routines (\textcolor{green}{\textbf{green color}}) enable
transactional processing of graph data.
Indexes (\textcolor{orange}{\textbf{orange color}}) provide indexing structures
for vertices and edges, speeding up different queries. Indexes heavily use
  constraints to provide indexing for vertices/edges satisfying specific
  conditions.  Routines for constraints are indicated with
  \textcolor{brick}{\textbf{brick red color}}.
Finally, all groups of routines heavily use error codes
(\textcolor{red}{\textbf{red color}}) and schemes for basic datatypes
(\textcolor{gray}{\textbf{gray color}}).
We now elaborate on key GDI parts. 
%

There are two classes of GDI routines: \textbf{collective} (``[C]'') and \textbf{local} (``[L]'').
All processes actively participate in a collective routine (i.e., they all
explicitly call this routine), while only a single process actively
participates in a local routine (it can still passively involve
arbitrarily many other processes by accessing their memories).
Collective
communication has heavily been used in high-performance computing~\cite{chan2007collective,
hoefler2014energy}. Such communication routines, by actively involving all
participating processes, are more efficient than routines based on
point-to-point communication by facilitating various optimizations and advanced
communication algorithms~\cite{hoefler2008accurately, hoefler2014energy}. They
also foster portability and programmability~\cite{hoefler2008accurately} by
coming with well-defined semantics for the behavior of groups of processes.

\enlargeSQ

\subsection{High-Performance Transactions}

A transaction consists of a sequence of operations on graph data, and it must
guarantee Atomicity, Consistency, Isolation, and Durability (ACID). GDI poses
no restriction on how to ensure ACID.
GDI transactions support full CRUD functionality for vertices, edges, and their
associated labels and properties.
Accessing and modifying graph data is conducted only within a transaction body.
Any single process can be in arbitrarily many concurrent transactions.

\iftr
Atomicity ensures that the operations are treated as a single unit and either
all succeed or completely fail. Consistency ensures that before and after a
transaction, the database is always in a consistent state. Isolation ensures
that concurrent transactions behave as if they were run in some sequential
order. Durability ensures that the effects of a committed transaction remain in
the database even in the case of a system failure.
\fi


\textbf{Local (single process) transactions} are transactions that a single
process has started.  This type of a transaction is meant for graph operations
which touch only a small part of the graph. \textbf{Collective transactions}
are transactions which actively involve all processes; they are used to execute
large OLAP or OLSP queries.

%
\includegraphics[scale=0.2,trim=0 16 0 0]{excl.pdf} \textbf{\ul{Major Design
Choice \& Insight}}: \emph{Use
collective transactions, that involve all processes, for global OLAP/OLSP workloads.
This facilitates not only low latencies (as collectives are highly
tuned) but also programmability (as collectives have well-defined semantics).}


GDI also distinguishes \textbf{read transactions} from 
\textbf{write transactions}. This further facilitates high-performance
implementations, by providing opportunities for optimized read-only
transactions that can assume that no participating process modifies the data.

\iftr
GDI differentiates between \textbf{transaction critical} and
\textbf{transaction non critical errors}. If a function returns a
transaction critical error, the transaction is guaranteed to fail. GDI does not
offer functions to retry a transaction or to recover from a
transaction critical error: The user must start a new transaction.
\fi

\enlargeSQ

\subsection{Fast \& Effective Access to Graph Data}
\label{sec:gdi-access}

GDI provides fast transactional access to vertices, and their labels and
properties, with a {two-step} scheme. 
In the first step, an application-level vertex ID is translated to an internal
GDI-specific ID. This makes GDI more portable, as it is independent of any
details of how the higher-level system layers may implement IDs.  In our
implementation, we use internal indexing structures, detailed in
Section~\ref{sec:gdi-impl}, for this translation. This internal ID uniquely
identifies a vertex in the whole GDI database.

GDI offers two types of internal IDs: \emph{volatile} and \emph{permanent}.
The former are valid only during the transaction within
which they are obtained. This facilitates optimizations such as the dynamic
relocation of graph data, but it also requires re-obtaining these IDs
in each transaction. The latter are shared across transactions,
which reduces the number of remote operations, but 
hinders dynamic load balancing. The user can choose the most
suitable variant.

\includegraphics[scale=0.2,trim=0 16 0 0]{excl.pdf} \textbf{\ul{Major Design
Choice \& Insight}}: 
\emph{Volatile IDs facilitate optimizations related to load balancing}. 
For example, it facilitates redistributing the graph across processes between
collective transactions, without fearing that internal IDs become stale.

\enlargeSQ

\subsection{Handles}

Internal representations of objects involved in transactions, such as
vertices or property types, are not directly accessible to the GDI user. 
To enable fast and programmable way of accessing and manipulating
graph data within transactions, GDI prescribes using \textbf{handles}
(\textbf{access objects}), i.e., opaque objects that hide the internal
implementation details of accessed objects, and represent these objects
\emph{on the executing process}. 
\iftr
These handles can be
passed as arguments, and they can participate in assignments and comparisons. 
Each vertex and edge handle must be
instantiated with an \textbf{association}.
\fi
To create a handle for an existing vertex $v$ or
edge $e$, the user calls \texttt{GDI\_AssociateVertex($v$)} or
\texttt{GDI\_AssociateEdge($e$)}; $v$ and $e$ are respective internal IDs. 
\iftr
Opaque objects and their handles are process specific, and they are only
significant at their respective allocating process and cannot be shared with
other processes via communication.
\fi

\includegraphics[scale=0.2,trim=0 16 0 0]{excl.pdf} \textbf{\ul{Major Design
Choice \& Insight}}: 
\emph{Using handles to access opaque objects improves
usability.} First, it enables the GDI implementation to decide on the details
of how graph data is accessed. Using a handle enables
remote direct (zero-copy) memory access, but it could also be used to
transparently copy or move the data, for example for dynamic relocation or to
cache the data locally.
Moreover, it relieves the user of ensuring that there are no pending operations
involving out-of-scope, opaque objects; the GDI implementation instead takes
care of that. It also allows users to simply mark objects for deallocation,
relying on the GDI implementation to retain the object until all pending
operations have completed. Requiring handles to support native-language
assignment and comparison operations keep the GDI interface clean and simple.

\iftr
GDI intends the allocation and deallocation of objects to appear to the user as
if the information for those objects were copied, so that semantically opaque
  objects are separate from each other. However, this does not mean that GDI
  implementations may not employ optimizations such as references and reference
  counting, if they are hidden from the user.
\fi

\iftr

\subsection{Flexible Indexes \& Constraints}

It is desirable for a graph databases to enable a fast lookup of vertices and
edges \emph{satisfying certain application-specific conditions}, for example
having a specific label or a property. 
%
%
Here, GDI provides an interface for indexing structures that can be explicitly
used by the GDI user. We enable this interface to have flexible indexing
conditions, to support a wide variety of graph database workloads.
Explicit indexes are queried by using boolean formulas (referred to as
\textbf{constraints} in GDI) in disjunctive normal form (DNF). GDI constraints
support arbitrary conditions regarding labels and properties.
Explicit indexes impose no restrictions on what indexes the
implementation of GDI might use internally.

%
\subsection{Performance-Centric Syntax \& Semantics}

GDI fosters many optimizations by enabling the GDI user to
provide GDI with extensive (but optional) information about their setting. As
an example, we discuss property types.
First, one can inform GDI whether there can be at most one property entry of 
a given property type, on a given single vertex or edge, or whether there may
be multiple entries of this property type on one vertex/edge.
Next, one can inform GDI about the datatype of the elements of the property
value.  Finally, one can also explicitly specify whether the property type has
a certain size limitation, or if it has a fixed size.
%

\fi

\subsection{Consistency}

For performance reasons, GDI enables different consistency models. The
interface requires \textbf{serializability} for graph data (vertices, edges,
and their associated labels and properties). Generally, this data can only be
altered by transactions that ensure ACID. Second, GDI guarantees
\textbf{eventual consistency} for metadata (labels, property types) and for
indexes. Since these objects also affect the graph data, this might lead to
cases where graph data becomes inconsistent until the system has converged.
Transactions must be able to detect such state and abort accordingly. Note that
implementations are free to provide consistency models for metadata and for
indexes that are more restrictive (stronger) than eventual consistency.

\includegraphics[scale=0.2,trim=0 16 0 0]{excl.pdf} \textbf{\ul{Major Design
Choice \& Insight}}: 
\emph{Enabling separate consistency models for data and metadata fosters
flexibility and simplicity.}

Many systems only specify their compliance with the Consistency requirement of
ACID, but do not clearly define what type of consistency they employ~\cite{neo4j_book}.
In GDI, we clearly specify it.

\includegraphics[scale=0.2,trim=0 16 0 0]{excl.pdf} \textbf{\ul{Major Design
Choice \& Insight}}: 
\emph{Clearly specifying the used consistency model fosters programmability.}
%

\iftr

\subsection{Multiple Parallel Databases}

GDI supports running multiple concurrent distributed GDBs in a
single environment. For this, GDI functions accept an optional handle to a
graph database object. While we do not currently focus on this scenario, we
decided to include it, predicting that it will facilitate future integration of
GDI with cloud settings where multiple users run their separate databases.

\fi

\section{GRAPH WORKLOADS WITH GDI}
\label{sec:workloads}


We now illustrate how to use GDI to easily and portably
implement representative queries from all major classes of GDB 
workloads.
In principle, one could implement all of these workloads with single-process
transactions. In GDI, we observe that, for some of these workloads,
if they harness all processes in a database, this gives more performance. Thus, it is
often more beneficial to use collective transactions in such cases.
We summarize what types of transactions are best to be used for what workloads 
in Table~\ref{tab:gdi-workload-support}.
\iftr
Note that these are general recommendations; {GDI users are free
to use any transaction type for any workload, depending on their domain
knowledge and insights}.
\fi

\begin{table}[h]
\centering
\setlength{\tabcolsep}{2.5pt}
\footnotesize
\begin{tabular}{llll@{}}
\toprule
\textbf{Workload class} & \multicolumn{2}{c}{\textbf{Type}} & \textbf{Best-suited GDI routines} \\
\midrule
Interactive (short) & read-only & OLTP & Single-process transactions \\
Interactive (complex) & read-only & OLTP & Single-process transactions \\
Interactive (updates) & read/write & OLTP & Single-process transactions \\
Graph analytics & read-only & OLAP & Collective transactions \\
Business intelligence & read-only & OLSP & Single-process or collective trans. \\
Massive data ingestion & read/write & BULK & Bulk data loading collectives \\
\bottomrule
\end{tabular}
\caption{Key graph database workloads (see Section~\ref{sec:back} for details)
and the associated recommended mechanisms of GDI best used for implementation.}
\vspaceSQ{-1em}
\label{tab:gdi-workload-support}
\vspaceSQ{-2em}
\end{table}

\iftr
We show how to use GDI to develop the above-mentioned workloads. 
\fi
\if 0
Listings~\ref{lst:trans_neigbor}, \ref{lst:gcn}, \ref{lst:bfs_traversal},
and~\ref{lst:business_workload} contain -- respectively -- a simple
interactive query (fetching properties from a small vertex set), two graph
analytics queries (a Graph Convolution Network (GCN) model 
from a recent class of Graph Neural Network (GNN) workloads, and a BFS
traversal), and a complex business intelligence transaction.
\else
Listings~\ref{lst:trans_neigbor}, \ref{lst:gcn},
and~\ref{lst:business_workload} contain -- respectively -- a simple OLTP
interactive query (fetching properties from a small vertex set), an OLAP query
(a convolutional Graph Neural Network (GNN)), and an OLSP
transaction.
\fi
For clarity, we omit straightforward additions (e.g., error handling or
checking if transactions fail).
In all the
queries, for each accessed vertex or edge, one first translates the
application-level ID to the GDI ID, and then uses the obtained ID to
create handles to be able to access the corresponding graph data. 
The used symbols are as follows: \texttt{trans\_obj} (a handle to the state of
the ongoing transaction), \texttt{vH} (a handle to a vertex $v$), \texttt{eH}
(a handle to an edge $e$), \texttt{vID} (an internal GDI ID for a vertex $v$),
\texttt{vID\_app} (an external application-level ID for a vertex $v$).

\begin{figure}[h!]
\centering
\begin{lstlisting}[language=C,caption=\textmd{C-style pseudocode of an example
interactive OLTP query with GDI.  Here\mbox{,} we retrieve the first and last
name of all persons that a given person\mbox{,} modeled with a vertex
vID\_app\mbox{,} is friends with. For this\mbox{,} we first obtain all edges of
vID\_app (line 4)\mbox{,} iterate over them to find edges corresponding to
friendships (lines 5-10)\mbox{,} preserve the corresponding neighbors (line 10)\mbox{,} and retrieve
the names and surnames of each such neighbor (lines 11-15).}, label=lst:trans_neigbor]
GDI_StartTransaction(&trans_obj);
GDI_TranslateVertexID(&vID, GDI_LABEL_PERSON, &vID_app, trans_obj); //Find internal vertex ID (vID) based on the application-level ID (vID_app)
GDI_AssociateVertex(vID, trans_obj, &vH); //Create a temporary access object for vertex vID
GDI_GetEdgesOfVertex(&eIDs, GDI_EDGE_UNDIRECTED, vH); //Retrieve all undirected edges
for each eID in eIDs do {
 GDI_AssociateEdge(eID, trans_obj, &eH); //Create a temporary access object for edge eID
 GDI_GetAllLabelsOfEdge(&labels, eH);
 if(/* one of the labels equals GDI_LABEL_FRIENDOF */) {
  GDI_GetVerticesOfEdge(&v_originID, &v_targetID, eH); //Retrieve target vertex
  neighborsID.add(v_targetID) /* add target vertex to neighborIDs data structure. Details of neighborsID are omitted for clarity */ } }
for each vID in neighborIDs do {
 GDI_AssociateVertex(vID, trans_obj, &vH);
 GDI_GetPropertiesOfVertex(&fName, GDI_PROP_TYPE_FNAME, vH);
 GDI_GetPropertiesOfVertex(&lName, GDI_PROP_TYPE_LNAME, vH);
 /* add fName, lName to the data structure to be returned */ }
GDI_CloseTransaction(&trans_obj);
\end{lstlisting}
\end{figure}

\begin{figure}[h!]
\centering
\iftr
\begin{lstlisting}[language=C,caption=\textmd{C-style pseudocode of an example OLAP query with GDI
(graph convolution network training/inference). The details of graph convolution are beyond the scope of this
work and they can be discussed in detail in rich existing
literature~\cite{wu2020comprehensive, besta2022parallel}. In brief\mbox{,} this query
consists of a specified number of iterations (``layers''). In each
layer\mbox{,} every vertex first updates itself based
on the features of its neighbors (``aggregation''\mbox{,} lines 9-12) and then the outcomes
are processed by a multilayer perceptron (MLP\mbox{,} line 13) and a non-linearity (line 14).
Finally\mbox{,} the property modeling the feature vector of each vertex is updated accordingly (line 15).
For clarity, we present a simplified query with the most important communication-intense
operations.},
label=lst:gcn]
for(l = 0; l < layers /* a user parameter */; ++l) {
 /* some form of collective synchronization */
 GDI_StartTransaction(&trans_obj);
 GDI_GetLocalVerticesOfIndex(&vIDs, v_index, trans_obj); //Retrieve local vertices
 for each vID in vIDs do {
  GDI_AssociateVertex(vID, trans_obj, &vH);
  GDI_GetPropertiesOfVertex(&feature_vec, GDI_PROP_TYPE_FEATURE_VEC, vH); //Get the vertex feature vector stored as a property
  GDI_GetNeighborVerticesOfVertex(&nIDs, GDI_EDGE_OUTGOING, vH); //Retrieve neighborhood vertices
  for each nID in nIDs do {
   GDI_AssociateVertex(nID, trans_obj, &nH);
   GDI_GetPropertiesOfVertex(&feature_vec_n, GDI_PROP_TYPE_FEATURE_VEC, nH);
   feature_vec += feature_vec_n; /* Apply the aggregation GNN phase; in this example, we use a summation */ }
  feature_vec = MLP(feature_vec); //Apply the update GNN phase; in this example, we use a simple MLP transformation defined externally by the user
  feature_vec = sigma(feature_vec); //Apply the non-linearity defined by the user
  GDI_UpdatePropertyOfVertex(&feature_vec, GDI_PROP_TYPE_FEATURE_VEC, vH); }
 GDI_CloseTransaction(&trans_obj); }
\end{lstlisting}
\fi
\end{figure}

\begin{figure}[h!]
\centering
\iftr
\begin{lstlisting}[language=C,caption=\textmd{C-style pseudocode of an example
business intelligence workload with GDI (see explanation of the Cypher query
in~\ref{sec:cypher_query}):
``MATCH (per:Person) WHERE per.age>30 AND per-:OWN->vehicle(:Car) AND
vehicle.color = red RETURN count(per)''. Here\mbox{,} we first fetch all vertices modeling people
(using an index\mbox{,} line 4)\mbox{,} check whether each such person satisfies the specified criteria
(lines 5-16)\mbox{,} including age (lines 7-8)\mbox{,} car ownership (lines 9-14)\mbox{,}
and the car color (lines 15-16).},
label=lst:business_workload]
local_count = 0;
GDI_StartCollectiveTransaction(&trans_obj);
//Index_obj indexes all vertices with label GDI_LABEL_PERSON
GDI_GetLocalVerticesOfIndex(&vIDs, index_obj, trans_obj);
for each person in vIDs do {
 GDI_AssociateVertex(person, trans_obj, &vH);
 GDI_GetPropertiesOfVertex(&age, GDI_PROP_TYPE_AGE, vH);
 if(age <= 30) { continue; } //The condition is not met
 /* Define a constraint "cnstr" with a label condition "== GDI_LABEL_OWN" (to check for the act of owning) */
 GDI_GetNeighborVerticesOfVertex(&things, cnstr, GDI_EDGE_OUTGOING, vH); //Get neighbors satisfying cnstr
 for each object in things do {
  GDI_AssociateVertex(object, trans_obj, &vH);
  GDI_GetAllLabelsOfVertex(&labels, vH);
  if(/* no label equals GDI_LABEL_CAR */) { continue; }
  GDI_GetPropertiesOfVertex(&color, GDI_PROP_TYPE_COLOR, vH);
  if(color == red) { local_count++; } } }
GDI_CloseCollectiveTransaction(&trans_obj);
reduce(local_count);
\end{lstlisting}
\fi
\end{figure}

\begin{figure*}[t]
\centering
%
\includegraphics[width=1.0\textwidth]{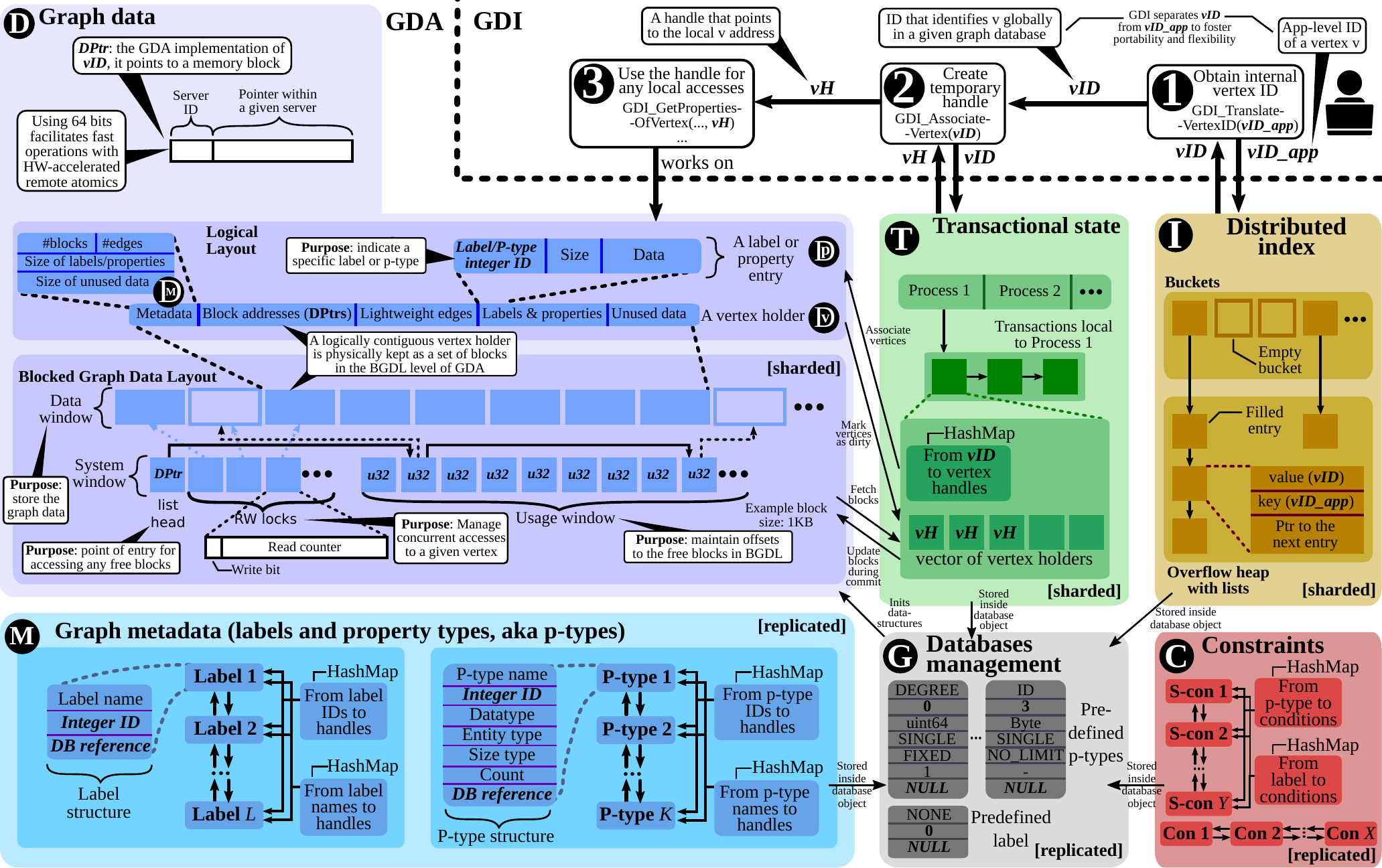}
\vspace{-2em}
\caption{Details of the GDI-RMA (GDA) implementation, and its interaction with GDI.
In the upper part of the figure, we illustrate a simple sequence of steps (taken
within a transaction) to access a selected property of a given vertex $v$.}
\vspaceSQ{-1em}
\label{fig:gda}
\end{figure*}

\section{SCALABLE GDI RDMA IMPLEMENTATION}
\label{sec:gdi-impl}

Our high-performance implementation of GDI, called GDI-RMA (GDA), is based on
MPI and it uses RDMA-enabled one-sided communication as the high performance
and high scalability driver.
\iftr

\subsection{RDMA \& RMA, and How We Use It}

RDMA has become widely used thanks to the high availability of RDMA-enabled
network interface cards (RNICs).
RDMA has been used with modern database management systems to accelerate data
replication~\cite{burke2021prism, jha2019derecho, kim2018hyperloop,
taleb2018tailwind, zamanian2019availability}, distributed
transactions~\cite{Wei_RDMA_HTM_transaction_paper, Dragojevic_farm_paper,
wei2018drtmh, Zamanian_rdma_transaction_paper}, distributed index
structures~\cite{ziegler2019rdmaindex}, general query
processing~\cite{Binnig_reldb_paper, Rodiger_reldb_1_paper}, and analytical
workloads~\cite{barthels2015rdmajoin, rdma_reldb_paper}.
\fi
While there are different ways to harness RDMA, \textbf{for highest
performance}, we focus on \textbf{one-sided fully-offloaded communication}.
Here, processes communicate by directly accessing dedicated portions of one
another's memories called a \emph{window}. Communication bypasses the OS and
the CPU, eliminating different overheads. Such accesses are conducted with
\emph{put}s and \emph{get}s that -- respectively -- write to and read from
remote memories. Puts/gets offer very low latencies, often outperforming
message passing~\cite{fompi-paper}. One can also use remote
\emph{atomics}~\cite{besta2015accelerating, schweizer2015evaluating, mpi3,
Herlihy:2008:AMP:1734069}; here, we additionally harness \textbf{hardware
support for atomics} offered by RDMA networks for very fast fine-grained
synchronization. For data consistency, we use \emph{flushes} to explicitly
synchronize memories. We use \textbf{non-blocking} variants of all functions,
because they can additionally increase performance by overlapping communication
and computation~\cite{fompi-paper}. All these routines are supported by virtually any
RDMA architecture, facilitating a wide portability of GDA.

\iftr
Fully offloaded one-sided RDMA usually outperforms two-sided communication both
in RDMA and in message passing. 
Yet, it also entails a complex and involved design, as conflicting accesses
must be fully resolved by sending processes. To alleviate this, we spent a
significant effort on ensuring that GDA is programmable, by making it highly
modular and easy to reason about. 
\fi

\iftr
\noindent
The semantics of the used operations are as follows: 

\enlargeSQ

\vspace{0.1cm}
\noindent {\ttfamily \footnotesize
GET(local, remote), PUT(local, remote),\\
CAS(local\_new, compare, result, remote)}
\vspace{0.1cm}

\noindent
where \texttt{GET} fetches a value from a remote address \texttt{remote} into a
local variable \texttt{local}, \texttt{PUT} insert the value of a local
variable \texttt{local} into a remote address \texttt{remote}, and \texttt{CAS}
compares a value at a remote address \texttt{remote}, and if it equals
\texttt{compare}, it replaces it with \texttt{local\_new}, saving the previous
value at \texttt{remote} into a local variable \texttt{result}). 
Atomic variants of put/get are indicated with \texttt{APUT/AGET}.
\fi

\iftr
\subsection{Overview of GDI-RMA Design}

We now overview GDI-RMA (GDA), see Figure~\ref{fig:gda}
(\includegraphics[scale=0.2,trim=0 16 0 0]{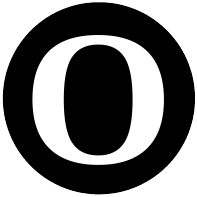}). 
\fi
We will detail the concepts
mentioned here in the following subsections.
Our implementation\footnote{\scriptsize We use the foMPI implementation of MPI
One-Sided routines~\cite{fompi-paper}.} is fully in-memory
for highest performance.
GDA consists of several modules which largely correspond to the GDI classes of
routines, cf.~Figure~\ref{fig:gdi-structure} (we use the same color code for
both illustrations).
The most important modules are management structures for GDI and parallel
databases (\includegraphics[scale=0.2,trim=0 16 0 0]{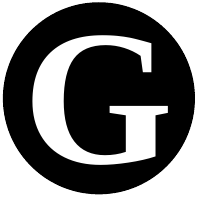}), metadata
structures (\includegraphics[scale=0.2,trim=0 16 0 0]{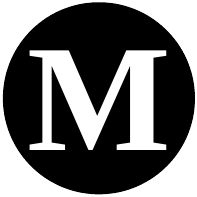}), indexes and
associated constraints (\includegraphics[scale=0.2,trim=0 16 0 0]{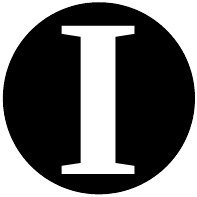},
\includegraphics[scale=0.2,trim=0 16 0 0]{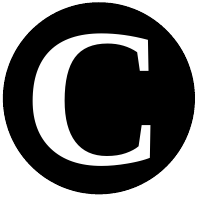}), state of collective and
local transactions (\includegraphics[scale=0.2,trim=0 16 0 0]{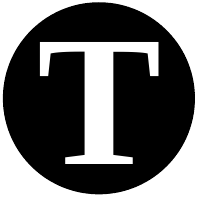},
and graph data
(\includegraphics[scale=0.2,trim=0 16 0 0]{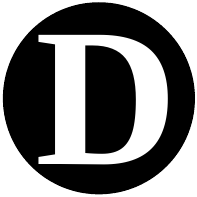}).
Structures that are small or independent of \#vertices and \#edges
(\includegraphics[scale=0.2,trim=0 16 0 0]{G.pdf},
\includegraphics[scale=0.2,trim=0 16 0 0]{M.pdf}, 
\includegraphics[scale=0.2,trim=0 16 0 0]{C.pdf}) are replicated on each
process to foster simplicity and high performance. All other structures are
sharded\footnote{\scriptsize Partitioning of data onto separate servers.}.
This, combined with our fully distributed transactions, enables fast
and scalable processing of very large graph datasets, being limited only by the
cluster size.

\enlargeSQ

\subsection{Graph Data}

The central part of GDA is associated with graph data
(\includegraphics[scale=0.2,trim=0 16 0 0]{D.pdf} in Figure~\ref{fig:gda}).
To make this part more manageable and programmable, it is divided into two
conceptual levels.
First, the \textbf{Logical Layout (LL)} level maintains structures that reflect
graph data (vertices, edges, labels, properties). Importantly, these
structures have \emph{flexible sizes determined by the sizes of the
corresponding parts of graph data}. For example, two vertices having different
sets of labels and properties would be maintained by two structures of
potentially different sizes.
The LL level simplifies working with GDI from the graph developer perspective,
because it enables a data-driven memory layout. However, it is challenging to
operate on such variable sized and dynamic structures in an RDMA environment
using one-sided communication.
For this, we also provide the underlying \textbf{Blocked Graph Data Layout
(BGDL)} level. BGDL maintains a large DM memory pool divided into same-sized
memory blocks (tunable by the user). The purpose of BGDL is to translate the
highly diverse structures from the LL level into these blocks. 
The memory blocks associated with one vertex/edge do not have to be stored
continuously, and might not even be located on the same process or server.
Such blocking enables a simple, effective, and flexible DM memory management:
any data access or manipulation routines operate on same-sized
blocks, and the difference between processing different parts of the graph is
only in the counts of the associated blocks.

\includegraphics[scale=0.2,trim=0 16 0 0]{excl.pdf} \textbf{\ul{Major Design
Choice \& Insight}}:
\emph{Introducing and separating the LL routines from the BGDL fosters
programmability. The LL routines form a clean and graph-centric API;
any performance optimizations can be done under the hood at the BGDL level.}

\iftr
\paragraph{Accessing Graph Data}

%
%
As an example, we discuss accessing a selected vertex~$v$ and its property,
see the top part of Figure~\ref{fig:gda}. 
\fi
Any access to the graph data
begins with the user providing an application ID (\textbf{\emph{vID\_app}}),
which is then translated to the internal ID \textbf{\emph{vID}} that uniquely
identifies a given object in the whole database, and can be shared by multiple
processes. This is conducted using the internal index
(\includegraphics[scale=0.2,trim=0 16 0 0]{I.pdf}).
In GDA, the internal ID is implemented as a 64-bit distributed hierarchical
pointer (\textbf{\emph{DPtr}}). Its first 16 bits indicate the compute server,
and the remaining 48 bits points to a local memory offset of the primary block
of $v$. We use 64 bits to facilitate using HW accelerated remote atomics, which
frequently operate on 64-bit words~\cite{fompi-paper}.
Then, \textbf{\emph{vID}} is used to construct a handle \textbf{\emph{vH}},
which is a pointer to $v$ in the local memory of a calling process. 
\iftr
Handles
are typed and the object they point to has to match that type.
Finally, while one can access any label or property of $v$ directly in $v$'s
holder, fetching additional information about the given metadata can be
done using label/property \emph{\textbf{integer IDs}} (used only within GDA).
\fi

\includegraphics[scale=0.2,trim=0 16 0 0]{excl.pdf} \textbf{\ul{Major Design
Choice \& Insight}}:
\emph{Using 64-bit distributed pointers facilitates harnessing hardware
accelerated remote atomic operations, which are commonly provided by
different vendors.}

\subsection{Logical Layout Level}

We shard the graph data (vertices, edges, and their associated labels and
properties) across all processes.
We implemented 1D (vertex-based) and 2D (edge-based) graph partitioning, and
use round-robin distribution (we tried other distribution schemes, they only
negligibly impact our performance).
GDI's specification is on purpose orthogonal to the partitioning/distribution,
so it is usable with any such scheme.

\subsubsection{Vertices \& Edges}

The data structure of each vertex~$v$ or an edge~$e$ (called a \emph{vertex} or
\emph{edge holder}, see \includegraphics[scale=0.2,trim=0 16 0 0]{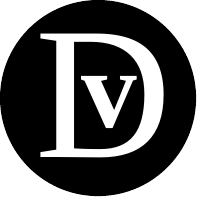}) is
divided into, respectively, metadata (selected important information used for
the data management, see \includegraphics[scale=0.2,trim=0 16 0 0]{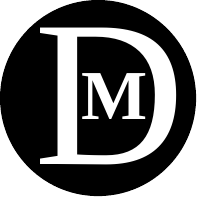}),
block addresses (addresses of blocks that store the data), lightweight edges
($v$'s edges that do not contain many labels or properties and are thus stored
together with $v$ for more performance), the label and property data (see
\includegraphics[scale=0.2,trim=0 16 0 0]{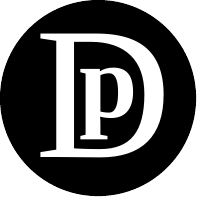}), and any unused memory.

\subsubsection{Lightweight Edges}

Many GDB queries involve iterating over edges
of a given vertex.  Simultaneously, in many graph datasets, only vertices have
rich additional data (labels, properties), while edges often do not carry such
data (e.g., in many citation networks). To maximize performance for these
cases, we introduce \emph{lightweight edges} in GDA.
Each such edge has at most one label. Importantly, these edges are stored in
the vertex holder object of their source vertex. This enables very fast
access. 

%
An edge UID is a data structure that identifies a lightweight edge uniquely in
the database. The data structure consists of two parts: a vertex UID and an
offset for the edge data structure of the vertex identified by the vertex UID.
Therefore the same edge can be identified by two different edge UIDs, depending
on which vertex is used as a base. An edge UID takes up 12 bytes of memory,
where the lower 8 bytes contain the vertex UID and the upper 4 bytes the offset
to the edge. 
%

\subsubsection{Labels \& Properties}

GDA uses a data structure called a \emph{label} or \emph{property entry} to store a single label or
property. GDA treats labels internally as properties for storage.
The \emph{\textbf{integer ID}} serves two purposes: it indicates whether an
entry is unused/empty (value 0) or whether it is the last entry (value 1),
and to store the integer ID of a given label/p-type (value 2 for a label,
any other value for a specific p-type).

\iftr
\subsection{Blocked Graph Data Layout (BGDL) Level}

The graph data (vertices and edges with their labels and properties) is mapped
to fixed-size blocks. 
The block size is specified by the user, enabling a
tunable tradeoff between communication amount and memory
consumption.
Specifically, the larger the blocks are, the less communication may be needed
to conduct DM graph operations (because fewer blocks are required to cover
given graph data), but also the more memory may be used, as larger parts
of blocks may be unavailable (as in internal fragmentation).
This tradeoff can be tuned by the user separately for different applications.
If the size of a memory block is big enough to contain all four memory
segments, only one memory block is used. The first memory block of a vertex is
called the primary block.

\includegraphics[scale=0.2,trim=0 16 0 0]{excl.pdf} \textbf{\ul{Major Design
Choice \& Insight}}:
\emph{Using fixed-size blocks in BGDL does not only simplify the design,
but it also fosters higher performance.} This is because one only needs
a single remote operation to fetch the data of a vertex that fits in one
block.
\fi

\iftr
Handling blocks is done internally and is not accessible through the library
interface. The two basic operations are

\vspace{0.1cm}
\noindent {\ttfamily \footnotesize
void acquireBlock( int target\_rank, DPtr* dp ),\\
void releaseBlock( DPtr* dp ).}
\vspace{0.1cm}


\texttt{acquireBlock} tries to allocate a block on process \texttt{target\_rank}. 
\texttt{releaseBlock} frees
a specified block. 
The block layer is oblivious to the actual contents of the blocks.

We use windows for the RMA based implementation of blocks.  There are three
windows: the \emph{data} window, the \emph{usage} window, and the \emph{system}
window. 
The data window contains all the blocks that constitute the vertex and
edge holder objects. Management of the blocks is done with the usage and the
system windows.
The usage window is essentially an indexing structure pointing to free blocks
in the general window. It holds a linked list, where each link element contains
the offset for the next free block in the data window.
Finally, the system window contains the pointer to the first unused element in
the linked list of the usage window as well as the locks that correspond to the
blocks stored on this process.

In block management, for highest performance, we use non-blocking communication
and lock-free synchronization. We now sketch the \texttt{acquireBlock} routine,
\texttt{releaseBlock} is similar.
First, the head of the list maintained by the usage window is retrieved with a
get from the system window. Second, the index of the next free block in the
data window is obtained with another get from the element that list head points
to. Third, we use an atomic CAS to replace the index of the current free
block in the list head with the index of the next free block. If the operation
is successful, no other process obtained a free block from the target process
in the time between the first get operation in step 1 and the issued CAS. In
this case, a distributed pointer with the rank of the target process and the
free block offset is created and returned to the origin process. If the CAS is
unsuccessful, the whole operation restarts at step 2 with the information just
obtained from the CAS. If the get in step 1 retrieves a NULL element, a NULL
handle is returned to the origin process to indicate that the target process
has no free blocks.

Block operations are prone to the ABA problem, because the list head is used to
check whether anything in the list has changed, while trying to complete an
acquire or release operation.
GDA alleviates this with the established tagged pointer technique. 

\fi

\subsection{Transactions \& ACI}

Each transaction is represented by a state with any necessary
information (e.g., which dirty blocks must be written back into the distributed
graph storage when the transaction commits). 
\iftr
The state of a single process
transaction is stored on its associated process, while the state of a
collective transaction is replicated on each process for performance reasons.
All changes applied in a transaction are visible only locally. 
If the
transaction commits, data changes are written to the remote blocks and any
associated indexes.

In order to ensure fast starts, commits, and aborts of transactions,
the operations on blocks performed under the hood must be highly efficient.
This includes verifying whether a transaction has already fetched blocks
corresponding to a certain vertex or edge holder object, tracking of dirty
blocks to be updated or removed, and deallocation of blocks.

To achieve high performance of these operations, a database maintains a list of
transaction handles, with each handle holding the address of the state of a
given transaction.
Each such state holds two hashmaps to efficiently lookup handles of vertex and
edge holder objects associated with a given transaction, given the vertex/edge
internal IDs. Moreover, the transaction state also holds a list with pointers to any local
and remote blocks used in the given transaction. 
When committing a transaction, either
all dirty blocks are written back or none, meaning that no elements must be
removed from the according list. Hence, we use a vector to keep track of the dirty
blocks that need to be written back during a commit phase. This enables adding
a new block address in ${O}(1)$ amortized
time and destroying a whole vector in ${O}(1)$ time.
\fi
\iftr
A linked list would instead require ${O}(n)$ for deconstruction (amortized:
  ${O}(1)$). 
\fi

\iftr

\includegraphics[scale=0.2,trim=0 16 0 0]{excl.pdf} \textbf{\ul{Major Design
Choice \& Insight}}:
\emph{Fast intra-transaction block management is key for high-performance
transactions. It can be achieved with using both hashmaps and linked lists
for keeping track of any blocks used within a transaction.}

\fi

\iftr
The user does not directly touch any block management. Instead, any data access
and modifications are conducted via with vertex and edge holder objects. These
objects and their associated routines handle the details of data caching and
fetching (e.g., to prevent double fetching of the same blocks).
\fi

\iftr
Similarly to the graph related data, we also keep track of the association
between each transaction and its associated graph database, to enable
concurrent database instances.
\fi

\iftr
\paragraph{Scalable Reader-Writer Locking for ACI}
\fi

GDA uses a two-phase scalable reader-writer (RW) locking to ensure the ACI properties. 
Only one lock per any vertex $v$ is used to reduce the number of remote
atomics. Figure~\ref{fig:gda} shows the lock data structure,
located in the system window at a corresponding offset to the primary block of
$v$'s holder object.
The \emph{write bit} determines if a process holds a write lock to $v$,
while the \emph{read counter} indicates the number of processes that currently hold a
read lock on $v$.
\iftr
The lock semantics are standard for RW locking, i.e., no write lock is in place
after the read lock is acquired, and no read locks are in place after the write
lock is acquired.
\fi

\enlargeSQ

\subsection{Lock-Free Internal Indexing}

GDA internally uses a distributed hashtable (DHT) to resolve different
performance-critical tasks conducted under the hood, such as mapping
application vertex IDs to internal GDA IDs. For highest performance and
scalability, GDA's DHT is \emph{fully-offloaded}, i.e., it only uses one-sided
communication, implemented with RDMA puts, gets, atomics, and flushes. Its
design is lock-free, it incorporates sharding, and it uses distributed chaining
for collision resolution.
To the best of our knowledge, this is the first DHT with all its operations being
fully offloaded, including deletes.
\ifconf
\else
The DHT consists of a table (to store the buckets) and a heap (to store linked
lists for chained elements). Each table entry consists of a key-value pair that
forms a bucket, followed by a distributed pointer to a linked list of entries
in the heap. Figure~\ref{fig:gda} (\includegraphics[scale=0.2,trim=0 16 0 0]{I.pdf}) shows the basic design of the DHT. 
The pseudocode is shown in Listing~\ref{lst:dhash_insert}.

\begin{figure}[h!]
\centering
\begin{lstlisting}[language=C,caption=\textmd{Distributed hash table operations used for internal indexing in GDA},label=lst:dhash_insert]
|\ul{insert}|(k, v) { // k: key, v: value
 bucket = hash(k); entry = alloc(); entry.k = k; entry.v = v;
 do {
  AGET(next_ptr, bucket); entry.next_ptr = next_ptr;
  CAS(&entry, next_ptr, result, bucket);
 } while(result != next_ptr); }

|\ul{lookup}|(k) { // k: key
 bucket = hash(k); AGET(next_ptr, bucket);
 if(next_ptr == NULL) return <false, NULL>;
 do {
  AGET(entry, next_ptr); // fetch the next ptr
  if(entry.next_ptr == next_ptr) {
   // entry is about to get deleted, so restart
   return lookup(k); }
  if(entry.k == k) return <true, entry.v>;
  next_ptr = entry.next_ptr;
 } while( next_ptr != NULL );
 return <false, NULL>; }

|\ul{delete}|(k) { // k: key
 bucket = hash(k); AGET(next_ptr, bucket);
 if(next_ptr == NULL) return false;
 // deleting the first entry is similar to the code below
 // prev_ptr is set correctly by now
 do {
  AGET(entry, next_ptr);
  if(entry.next_ptr == next_ptr) {
   // entry is about to get deleted, so restart
   return delete(k); }
  if(entry.k == k) {// found entry in question
   CAS(next_ptr, entry.next_ptr, result, next_ptr+2);
   if(result != entry.next_ptr) {
    // either entry is about to be deleted by someone else
    // or the next entry got deleted, so restart
    return delete(k); }
   CAS(entry.next_ptr, next_ptr, result, prev_ptr+2);
   if(result == next_ptr) { dealloc(next_ptr); return true; }
   else { // previous entry is about to be deleted
    // remember pointer to the next entry
    return delete(k, entry.next_ptr); } }
   prev_ptr = next_ptr; next_ptr = entry.next_ptr;
 } while(next_ptr != NULL);
 return false; }
\end{lstlisting}
\end{figure}

\textbf{Insert}
First, a distributed pointer is fetched from the bucket (line 4) and used as
the next pointer of the entry to be inserted. Then, an atomic CAS is used to
introduce the entry into the linked list (line 5). If CAS fails, then another
insert or delete happened concurrently, and the operation has to start again.

\textbf{Lookup}
In general, the operation follows the bucket linked list to find the entry with
the key in question. First, it fetches the distributed pointer from the bucket
(line 9), to see if there are any entries in that bucket (line 10). If there
are, each entry is fetched (line 12) and checked whether it contains the key in
question (line 16). If the next pointer points to itself, it indicates that the
entry is being deleted (line 13). In such a case, the lookup restarts (line 15).

\textbf{Delete}
The general loop is very similar to lookup, except that an additional
distributed pointer to the previous entry in the linked list is kept. The
difference between the two functions occurs, once the key in question is found.
Two CASes are necessary to delete a single entry. The first CAS changes the
next pointer of the entry to be deleted, so it points to the entry itself (line
32). This marks that entry as ready for deletion, in case of concurrent
lookups. If the 1st CAS fails (line 33), then either another process is
trying to delete the entry in question and has won that race, or the following
entry in the linked list was just deleted. In such a case, delete restarts
(line 36).
If the 1st CAS succeeds, the 2nd CAS tries to bypass the entry to be
deleted by pointing the next pointer of the previous entry to the succeeding
entry after the entry to be deleted (line 37). If the CAS succeeds, the entry
is deallocated (line 38) and the function returns. An unsuccessful
2nd CAS indicates that the previous entry is about be deleted as well, so we
restart, but retaining the original next pointer to the succeeding entry in the
linked list (line 41).

\fi

\includegraphics[scale=0.2,trim=0 16 0 0]{excl.pdf} \textbf{\ul{Major Design
Choice \& Insight}}:
\emph{Fully offloaded RDMA design facilitates high performance.
While being complex, it is kept under the hood and does not adversely impact 
GDI's programmability.}

\subsection{Graph Metadata}

We replicate graph metadata on each process for performance reasons. This is
because both $L$ and $P$ are in practice much smaller than $n$.
A {label} is represented by a structure that holds a label name, an integer ID,
and a reference to the associated graph database.
A property structure is similar, with the difference that it contains
additional information (the entity type, the GDI datatype, the size type, and
the size limitation).
We summarize these structures in Figure~\ref{fig:gda}
(\includegraphics[scale=0.2,trim=0 16 0 0]{M.pdf}).
When storing specific labels and properties on vertices/edges, we only use
their associated integer IDs.
To enable fast accesses to graph metadata, we maintain {double linked-lists} of
labels and properties, as well as hash maps.
The former enables to add and remove labels or properties in $O(1)$
work (given the handle), the latter is used for checking their existence in
$O(1)$.

\includegraphics[scale=0.2,trim=0 16 0 0]{excl.pdf} \textbf{\ul{Major Design
Choice \& Insight}}:
\emph{Replicating metadata simplifies the design without significantly increasing the
needed storage.}

\subsection{Summary of Parallel Performance Analysis}

\iftr
Each routine in GDA is supported with theoretical analysis of its performance,
in order to ensure GDA's \emph{performance portability}
(i.e., independence of
GDA's performance of various architecture details).
\fi
For this, we use the \textbf{work-depth (WD) analysis}. 
\iftr
The WD analysis enables
bounding run-times of parallel algorithms. 
\fi
Intuitively, the \emph{work} of a
given GDA routine is the total number of operations in the execution of this
routine, and the \emph{depth} is the longest sequential chain of dependencies
in this execution~\cite{Bilardi2011, blelloch2010parallel}.

Due to space constraints, we provide the work and depth of GDA routines in a
full extended version of the GDA manuscript.
Importantly, the majority of GDA routines (both for data and metadata
management) come with \emph{constant $O(1)$ work and depth}.
Only a few routines that modify $x$ metadata items (property types or labels)
come with $O(x)$ work and depth. This also implies \emph{low overheads in practice},
as $x$ is usually a small number (i.e., fewer than 10-20).

\iftr
\begin{figure*}[hbtp]
\centering
\vspaceSQ{-1.5em}
%
\begin{subfigure}[t]{0.48 \textwidth}
\centering
\includegraphics[width=\textwidth]{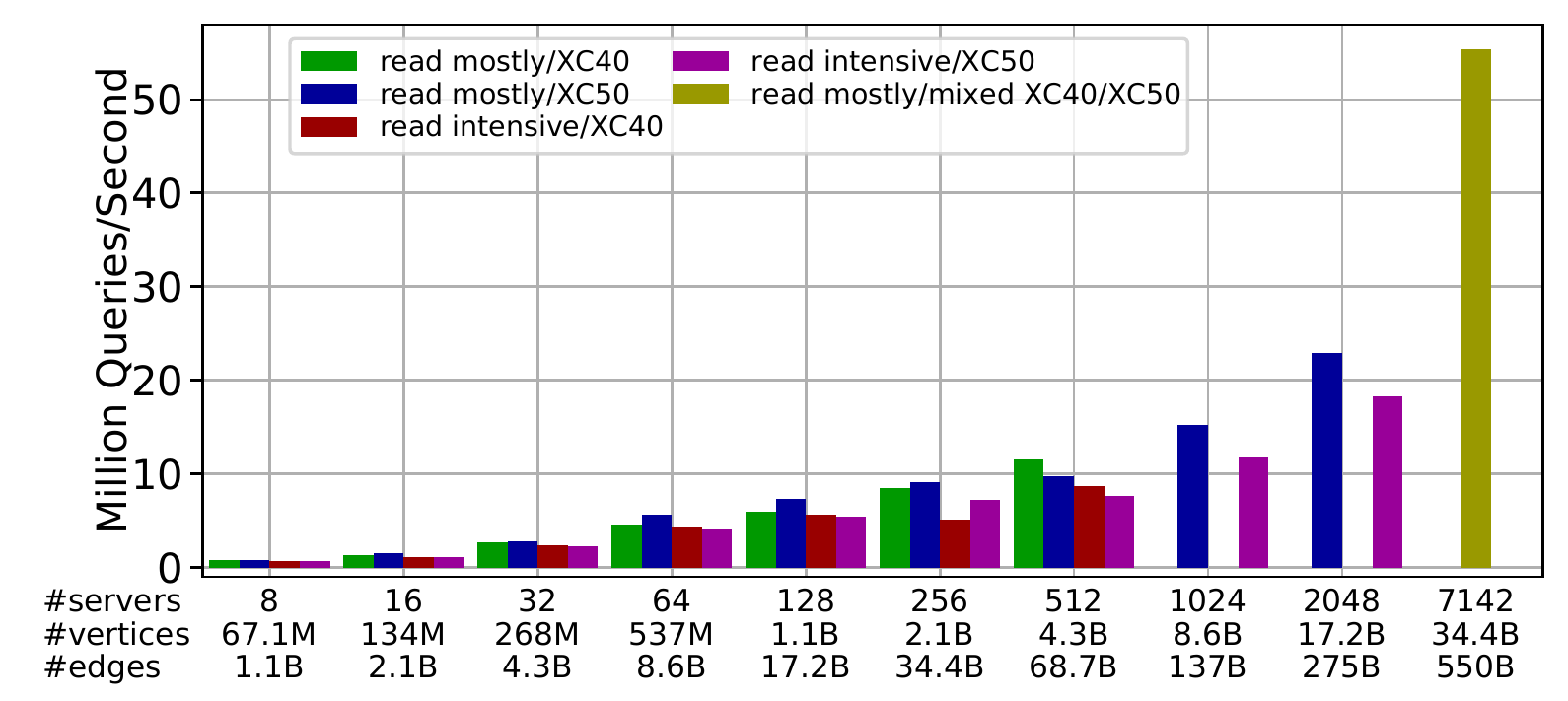}
\vspaceSQ{-2em}
\caption{\textmd{Read Intensive, Read Mostly; weak scaling.}}
\label{fig:oltp-read-weak-scaling}
\end{subfigure}
%
%
\begin{subfigure}[t]{0.48 \textwidth}
\centering
\includegraphics[width=\textwidth]{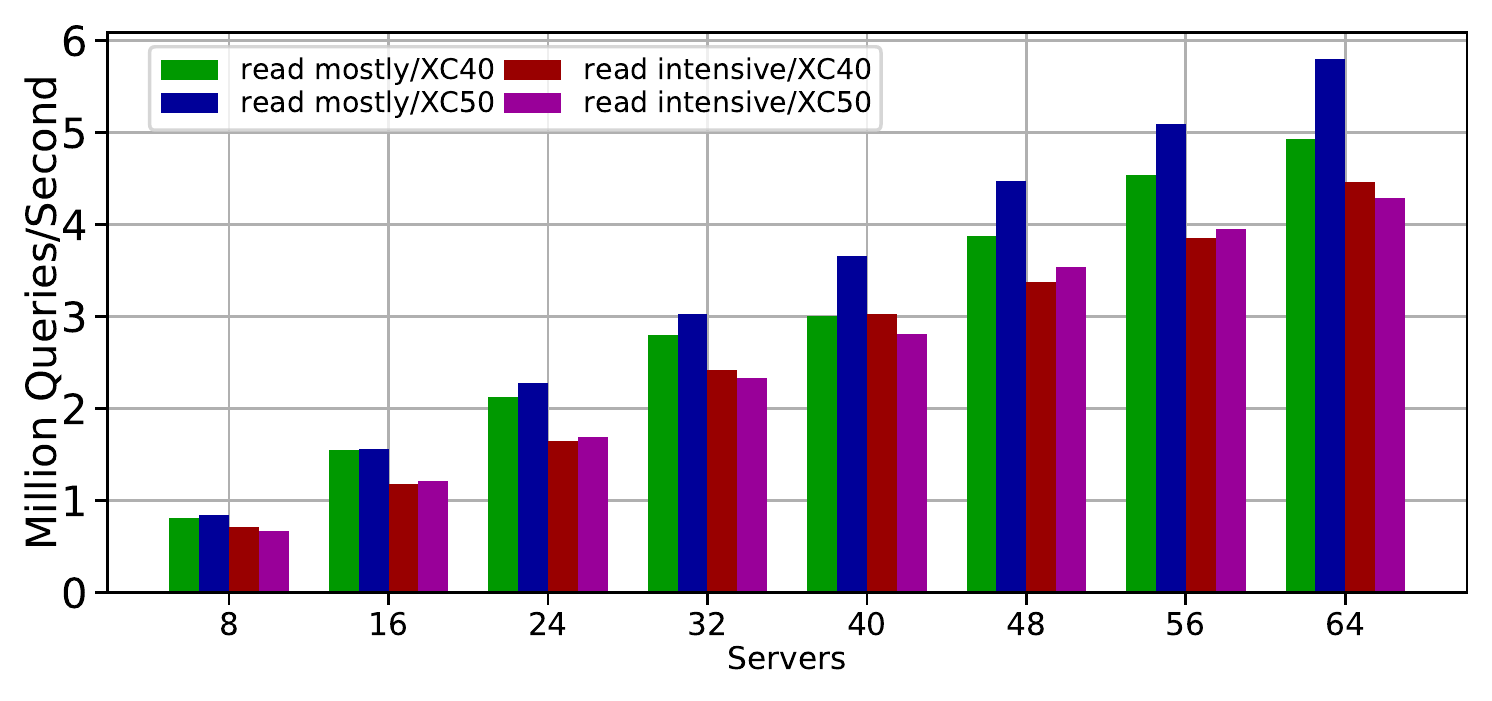}
\vspaceSQ{-2em}
\caption{\textmd{Read Intensive, Read Mostly; strong scaling.}}
\label{fig:oltp-read-strong-scaling}
\end{subfigure}
\begin{subfigure}[t]{0.48 \textwidth}
\centering
\includegraphics[width=\textwidth]{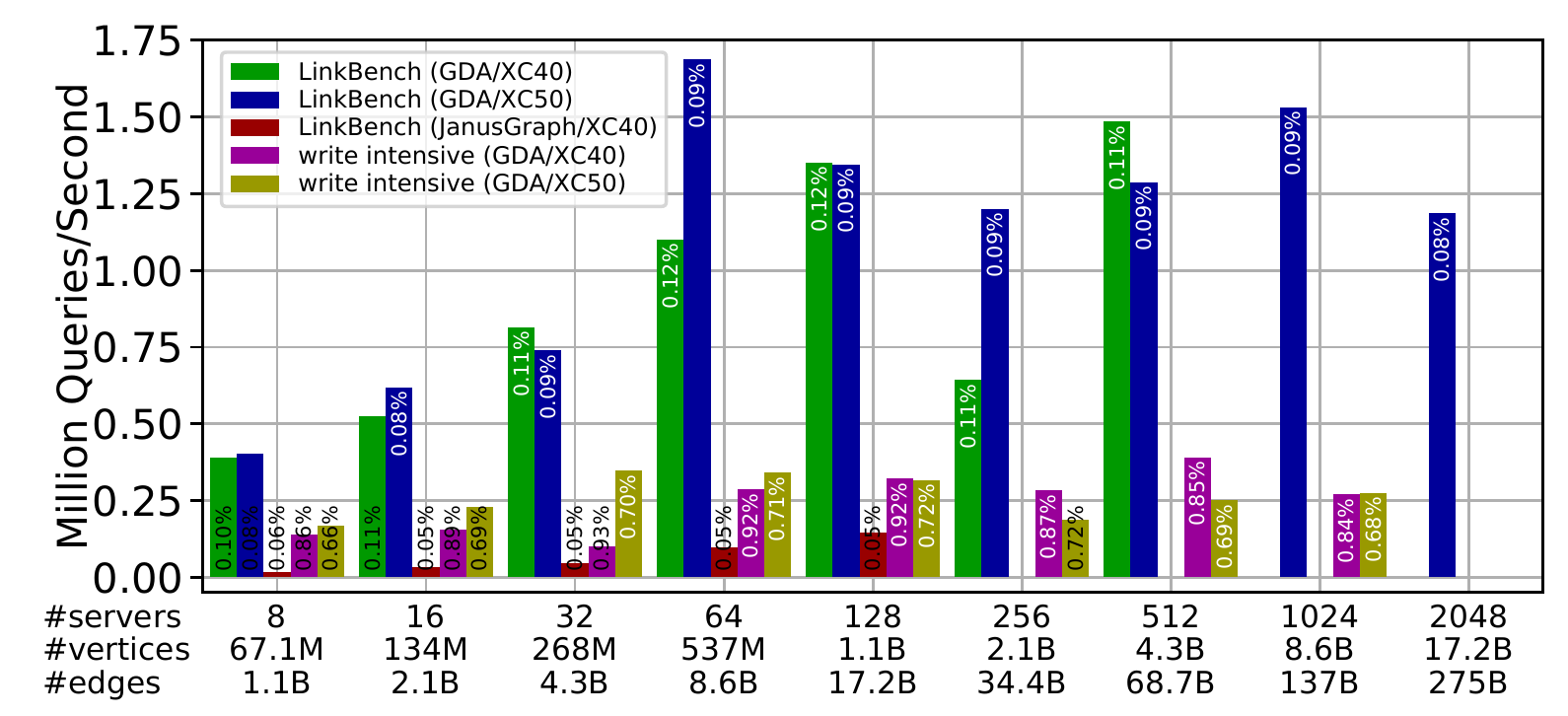}
\vspaceSQ{-2em}
\caption{\textmd{LinkBench, Write Intensive; weak scaling.}}
\label{fig:oltp-write-weak-scaling}
\end{subfigure}
%
%
\begin{subfigure}[t]{0.48 \textwidth}
\centering
\includegraphics[width=\textwidth]{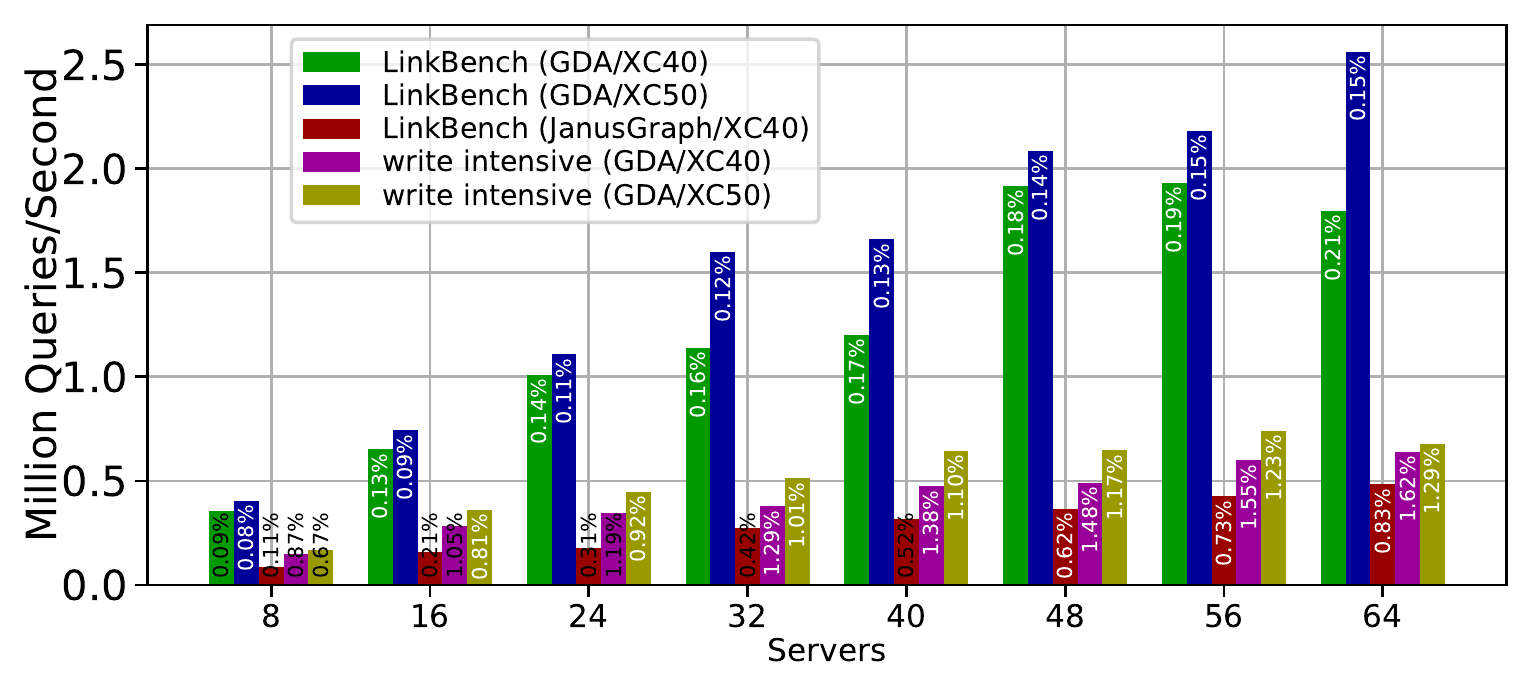}
\vspaceSQ{-2em}
\caption{\textmd{LinkBench, Write Intensive; strong scaling.}}
\label{fig:oltp-write-strong-scaling}
\end{subfigure}
\vspaceSQ{-1em}
\caption{\textmd{Analysis of OLTP workloads.
%
%
%
\textbf{XC40}, \textbf{XC50}: two types of servers considered
(cf.~Section~\ref{sec:setup}).  \textbf{Weak scaling}: scaling dataset sizes together with
\#servers, \textbf{strong scaling}: scaling \#servers for a fixed dataset
(a Kronecker graph of scale 26, i.e., 67.1M vertices and 1.1B edges; the results
follow the same performance patterns for other datasets).
\textbf{Missing bars} of our baselines indicate limited compute budget; missing
baselines of comparison targets indicate inability to scale to a given
configuration.
\textbf{Percentages}: the fractions of failed transactions (no percentage indicates
no, or negligibly few, failed transactions).}}
\label{fig:eval-oltp}
\end{figure*}
\fi

\section{EVALUATION}

\enlargeSQ

We now illustrate how GDI and its implementation GDA 
ensure high performance (latency, throughput) and large scale.

\subsection{Experimental Setup, Workloads, Metrics}
\label{sec:setup}

We first sketch the evaluation methodology. For measurements, we omit the first
1\% of performance data as warmup. We derive enough data for the mean and 95\%
non-parametric confidence intervals. We use arithmetic means as
summaries~\cite{hoefler2015scientific}.

As \textbf{computing architectures}, we use the Piz Daint Cray supercomputer
installed in the Swiss National Supercomputing Center (CSCS). Piz Daint hosts
1,813 XC40 and 5,704 XC50 servers. Each XC40 server has two Intel Xeon E5-2695
v4 @2.10GHz CPUs (2x18 cores, and 64 GB RAM). Each XC50 server comes with a
single 12-core Intel Xeon E5-2690 HT-enabled CPU, and 64 GB RAM).
The interconnect between servers is Cray's Aries based on the Dragonfly
topology~\cite{aries, dally08}.
We use \textbf{full parallelism}, i.e., we run
algorithms on the maximum number of cores available. 

\iftr
For evaluated \textbf{workloads}, we selected all three main classes of queries
as specified by the LDBC and LinkBench benchmarks~\cite{ldbc,
armstrong2013linkbench} and described in Section~\ref{sec:workloads}, namely
interactive queries~\cite{ldbc_snb_specification}, graph
analytics~\cite{ldbc_graphanalytics_paper}, and business intelligence
queries~\cite{early_ldbc_paper}.
\fi

We consider three \textbf{metrics}: \textbf{latency} (i.e., how fast a query
finishes), \textbf{throughput} (i.e., how many queries can we execute per time unit),
and \textbf{scale}. For scale, we (1) increase the number of
servers \emph{together with} the size of the dataset (the so called
``\textbf{weak scaling}'') and (2) increasing the number of servers \emph{for a fixed}
dataset (the so called ``\textbf{strong scaling}'').  
\iftr
The former evaluates how
GDA enables handling the increasing sizes of datasets, while the latter
illustrates how GDA is able to process the existing graphs faster.
\fi

\begin{figure*}[hbtp]
\centering
%
\begin{subfigure}[t]{0.32 \textwidth}
\centering
\includegraphics[width=\textwidth]{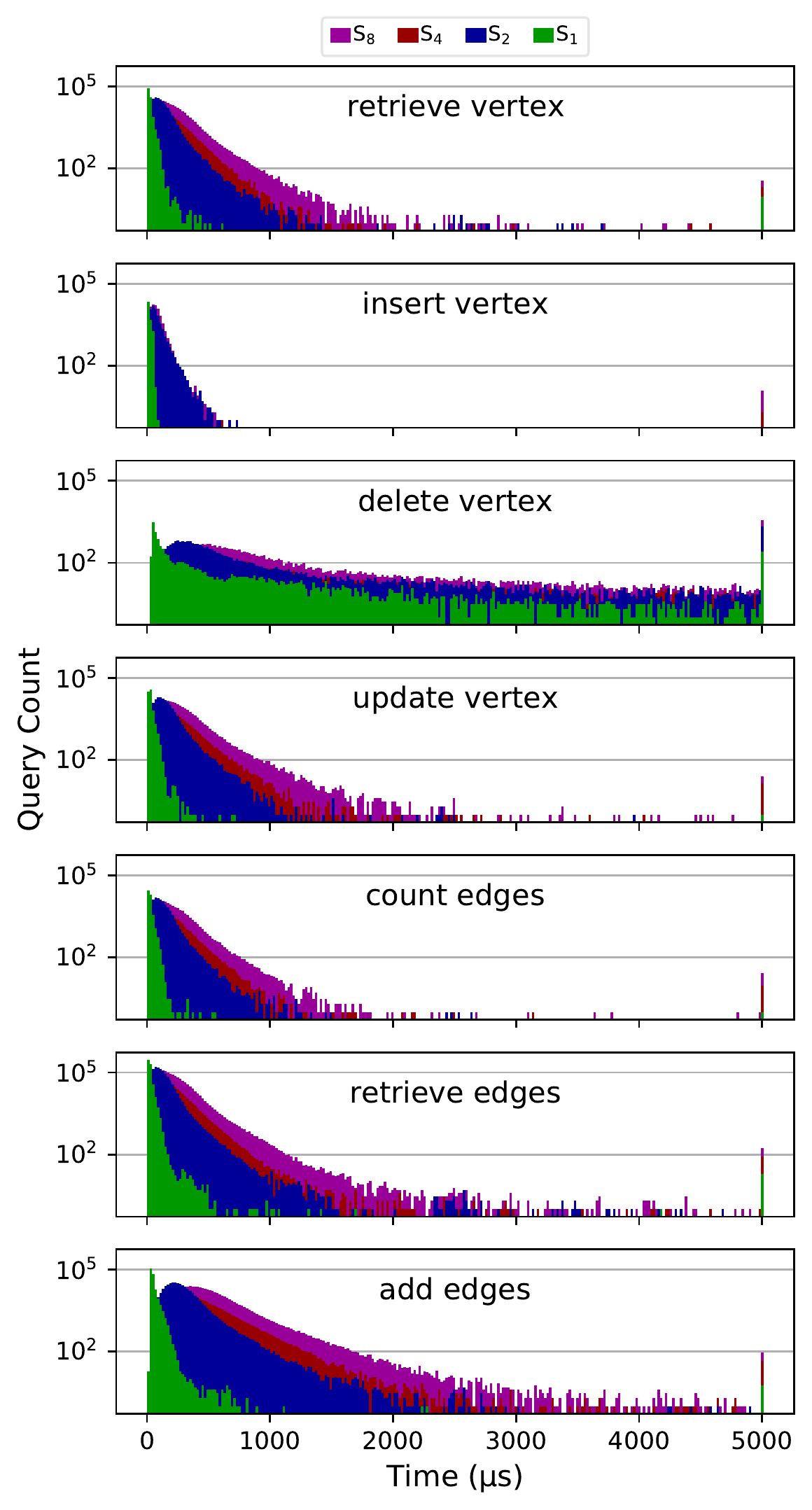}
\vspaceSQ{-1.5em}
\caption{\textmd{GDA.}}
\label{fig:oltp-gda-dets}
\end{subfigure}
%
%
\begin{subfigure}[t]{0.32 \textwidth}
\centering
\includegraphics[width=\textwidth]{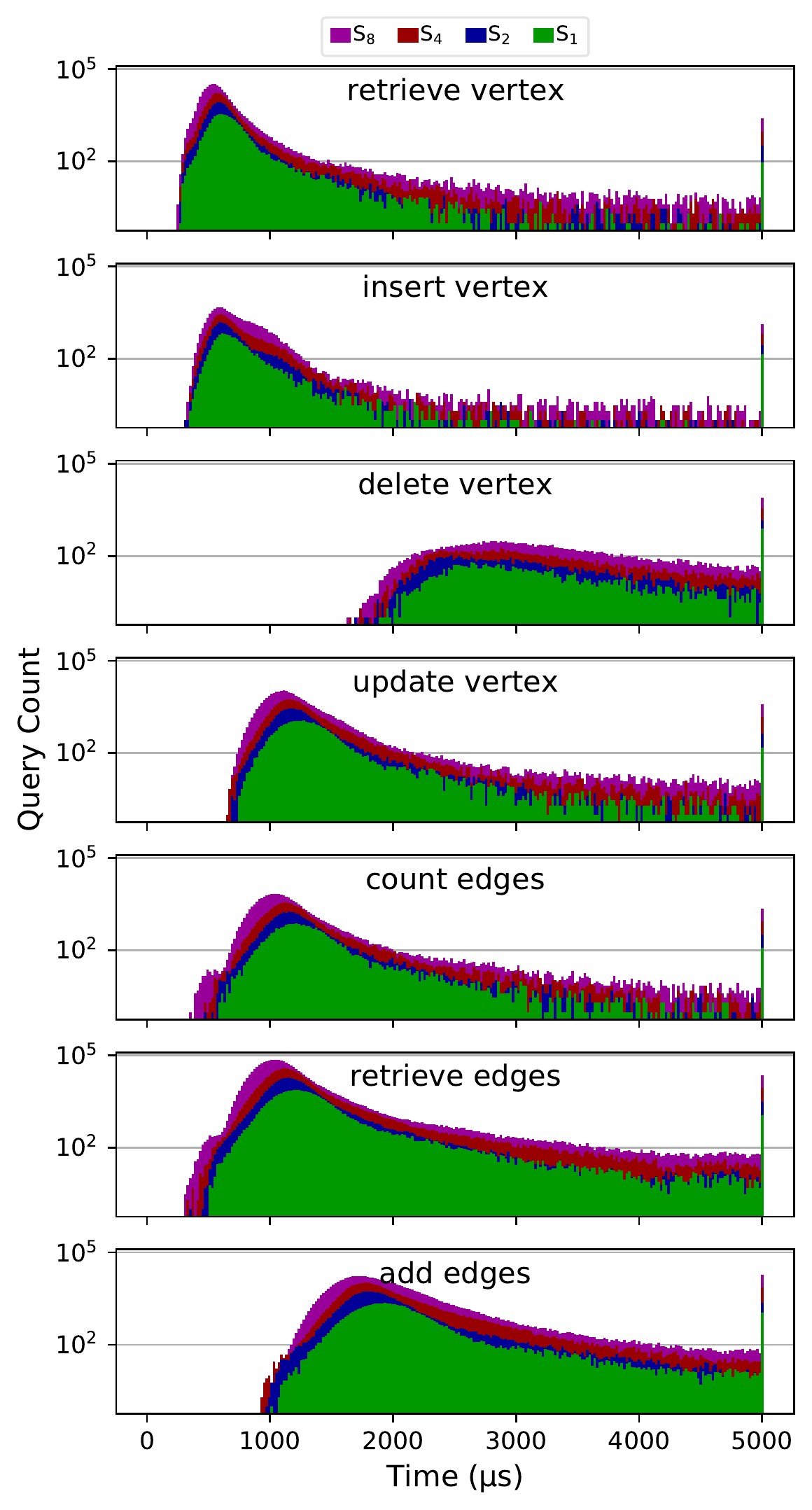}
\vspaceSQ{-1.5em}
\caption{\textmd{JanusGraph.}}
\label{fig:oltp-jg-dets}
\end{subfigure}
\begin{subfigure}[t]{0.32 \textwidth}
\centering
\includegraphics[width=\textwidth]{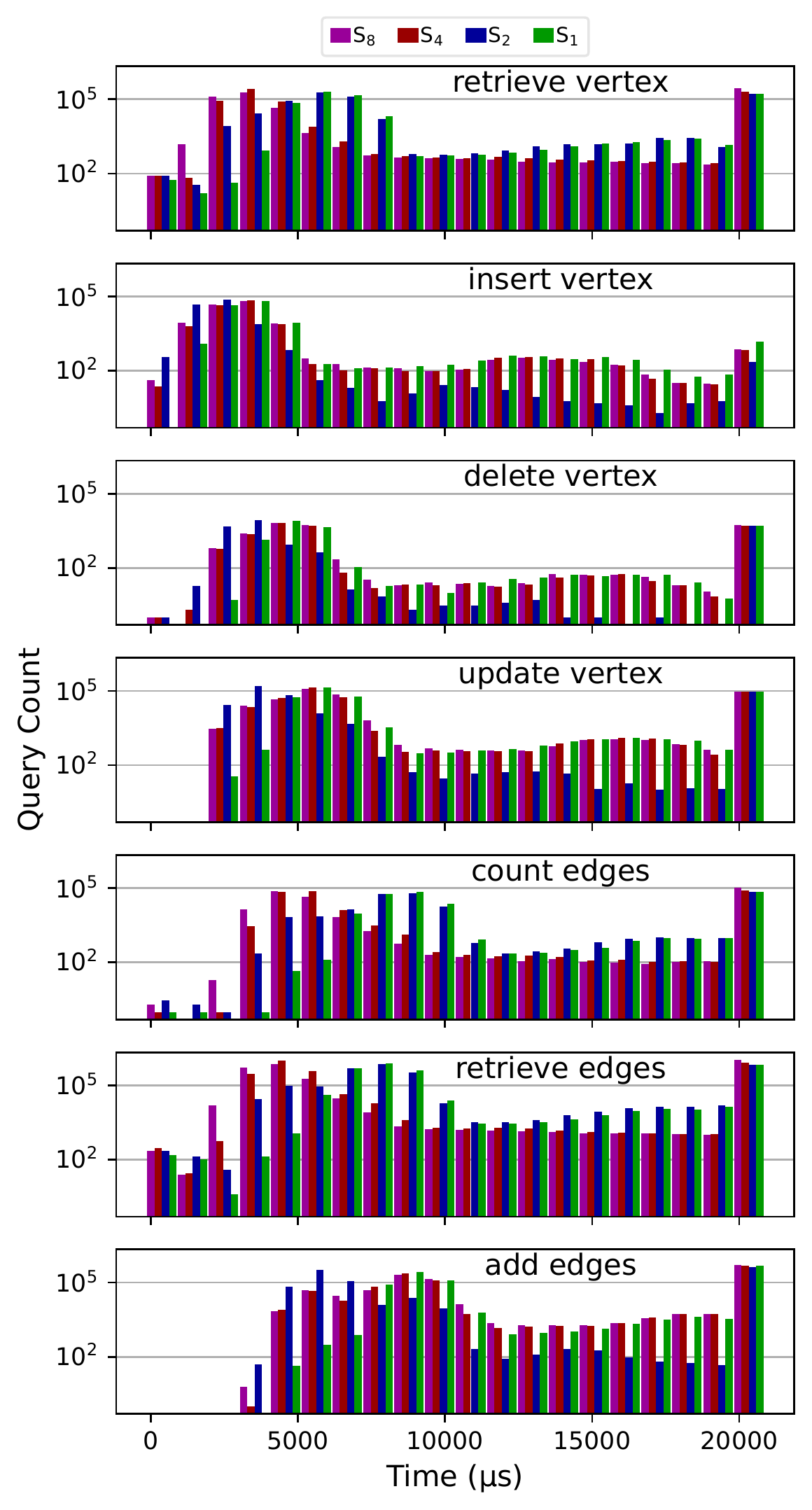}
\vspaceSQ{-1.5em}
\caption{\textmd{Neo4j.}}
\label{fig:oltp-neo4j-dets}
\end{subfigure}
\vspaceSQ{-0.5em}
\caption{\textmd{Details (histograms) on the latency of individual operations
of the OLTP LinkBench workload for 1, 2, 4 and 8 servers. We aggregate query
latencies outside the range and plot their combined number at the upper bound.
\textbf{Note the different cutoff of the X axis for Neo4j (20 ms) and GDA as well as
JanusGraph (5 ms)}. Because Neo4j only supports a coarser granularity of milliseconds
(versus $\mu$s for GDA and JanusGraph) for query
time measurements, we use a different cutoff and style for plotting.
S$_x$ indicates the data from a specific number of servers.}}
\label{fig:eval-oltp-dets}
\end{figure*}

\begin{figure*}[hbtp]
\centering
%
\begin{subfigure}[t]{0.49 \textwidth}
\centering
\includegraphics[width=\textwidth]{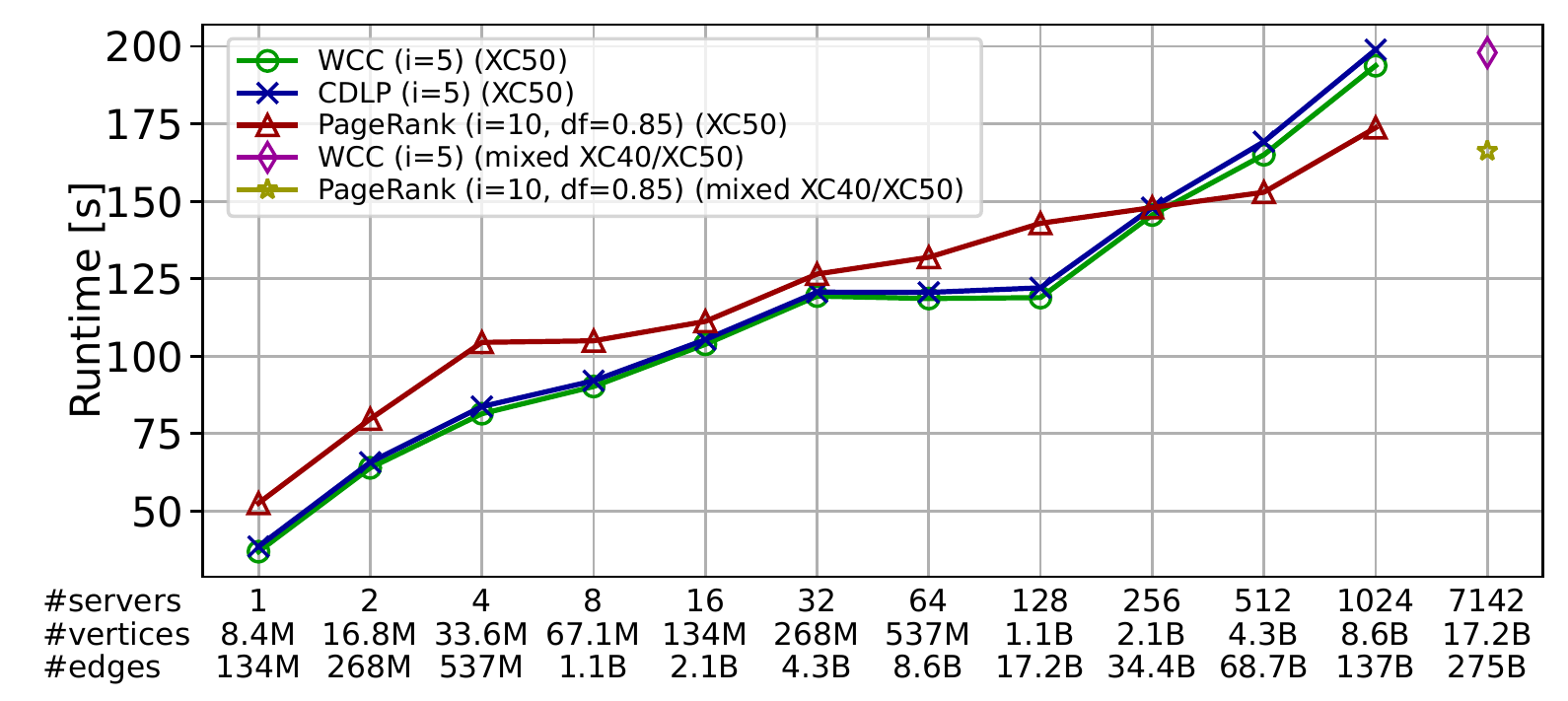}
\vspaceSQ{-1.5em}
\caption{\textmd{PR, CDLP, WCC; weak scaling.}}
\label{fig:olap-global-weak-scaling}
\end{subfigure}
%
%
\begin{subfigure}[t]{0.49 \textwidth}
\centering
\includegraphics[width=\textwidth]{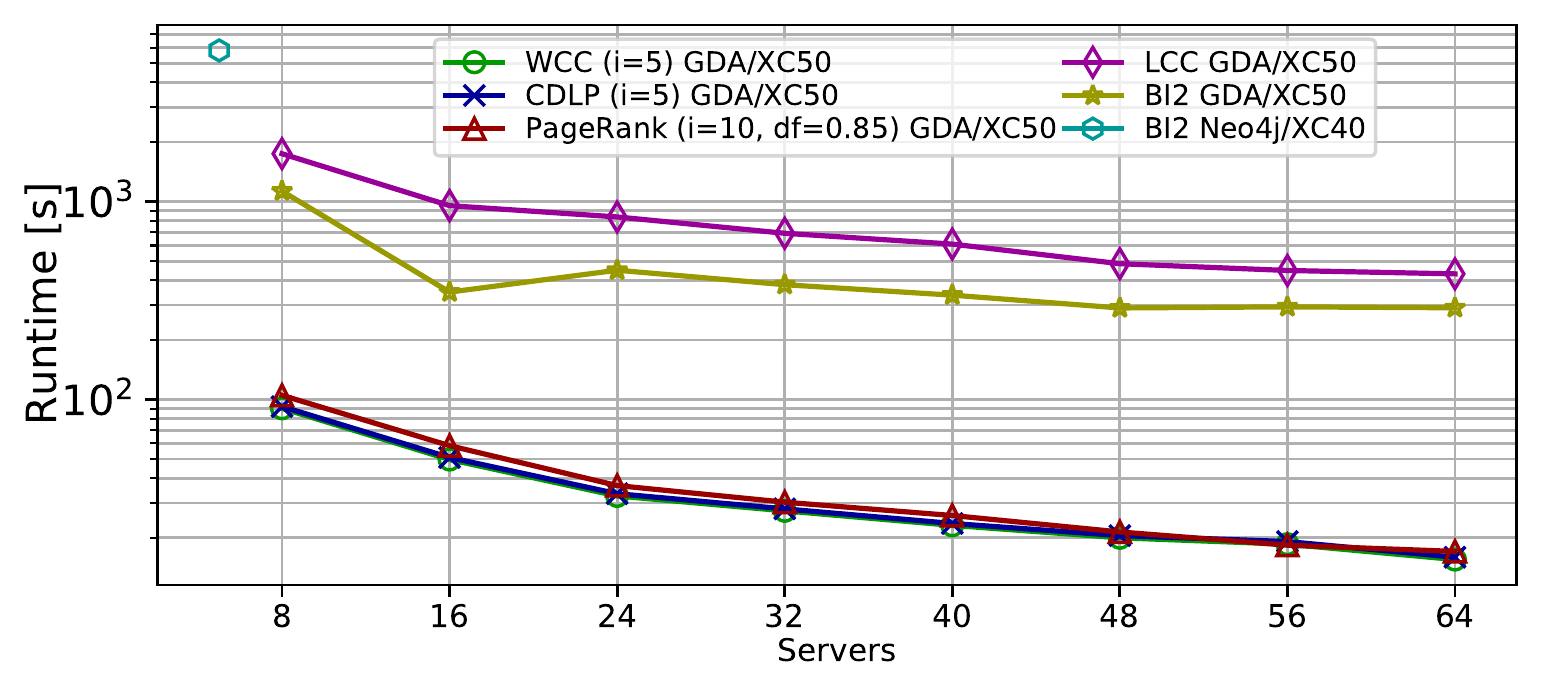}
\vspaceSQ{-1.5em}
\caption{\textmd{PR, CDLP, WCC, LCC, BI2; strong scaling.}}
\label{fig:olap-global-strong-scaling}
\end{subfigure}
%
%
%
\begin{subfigure}[t]{0.49 \textwidth}
\centering
\includegraphics[width=\textwidth]{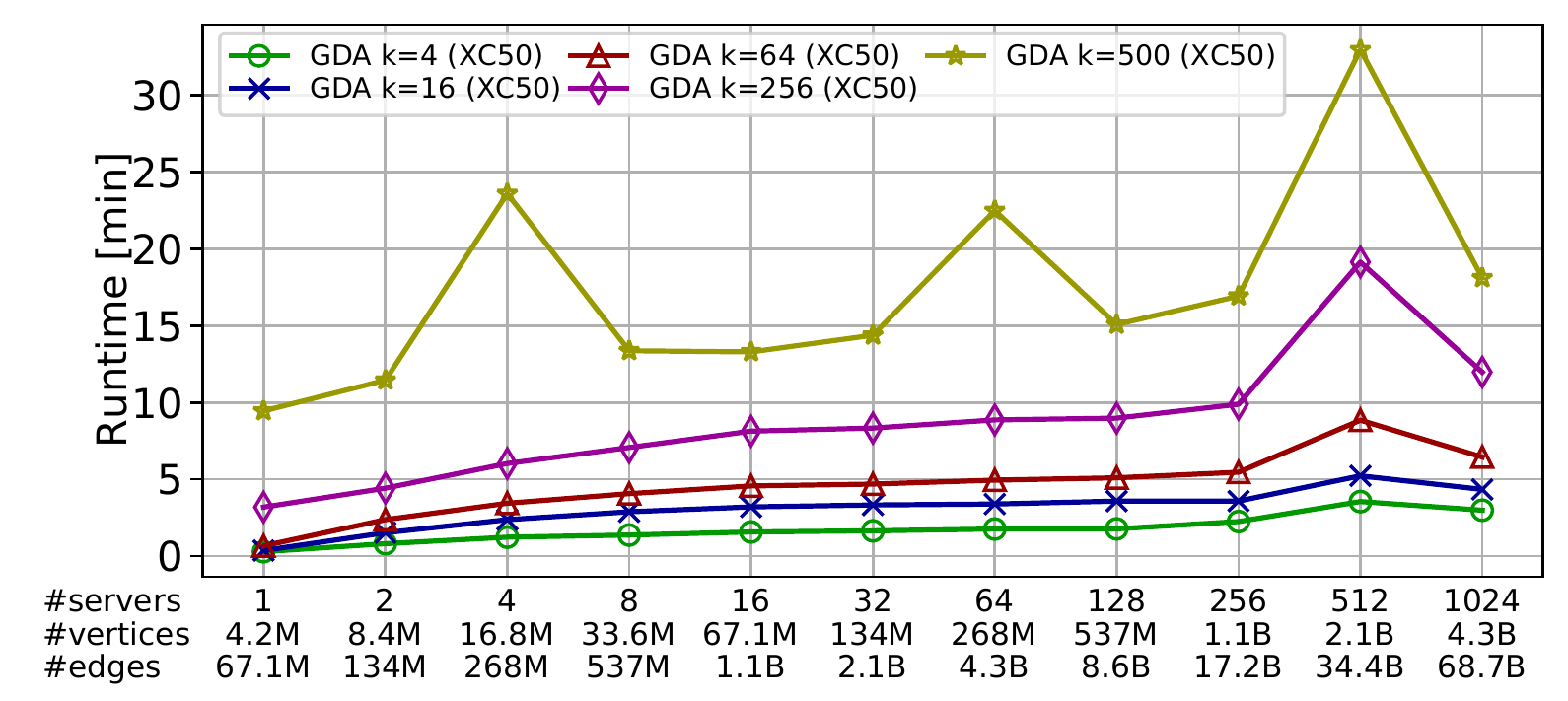}
\vspaceSQ{-1.5em}
\caption{\textmd{GNN; weak scaling.}}
\label{fig:olap-gnn-weak-scaling}
\end{subfigure}
%
%
%
\begin{subfigure}[t]{0.49 \textwidth}
\centering
\includegraphics[width=\textwidth]{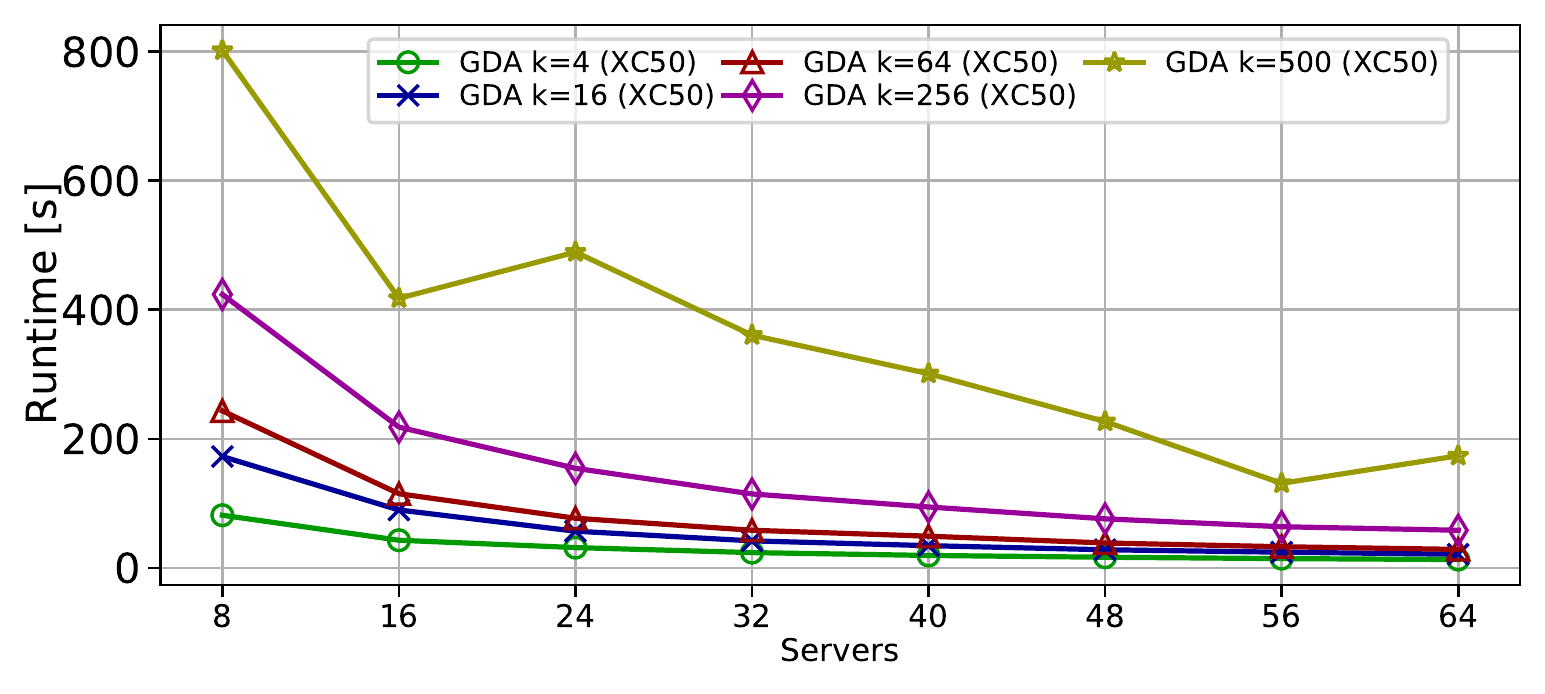}
\vspaceSQ{-1.5em}
\caption{\textmd{GNN; strong scaling.}}
\label{fig:olap-gnn-strong-scaling}
\end{subfigure}
%
%
%
\begin{subfigure}[t]{0.49 \textwidth}
\centering
\includegraphics[width=\textwidth]{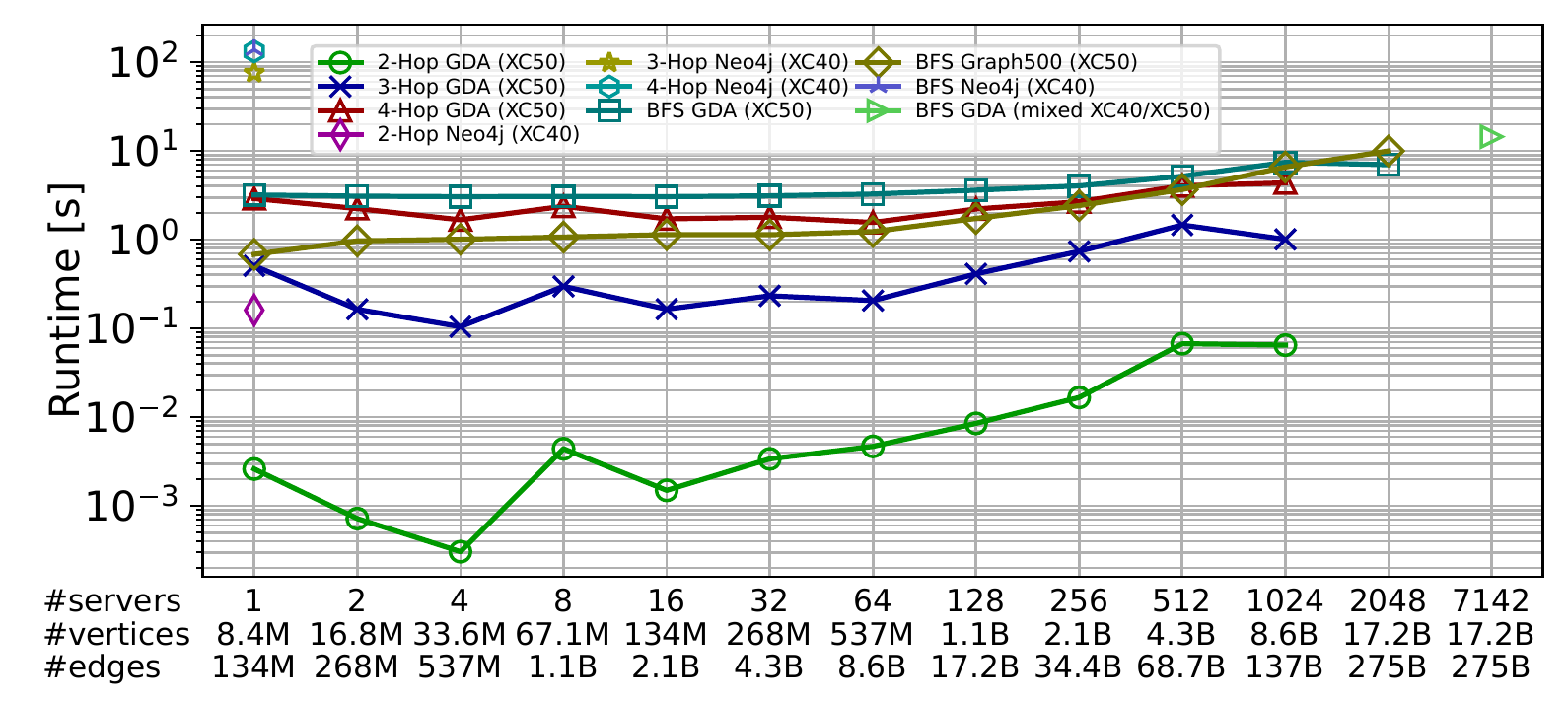}
\vspaceSQ{-1.5em}
\caption{\textmd{BFS, k-hop; weak scaling.}}
\label{fig:olap-bfs-khop-weak-scaling}
\end{subfigure}
%
%
%
\begin{subfigure}[t]{0.49 \textwidth}
\centering
\includegraphics[width=\textwidth]{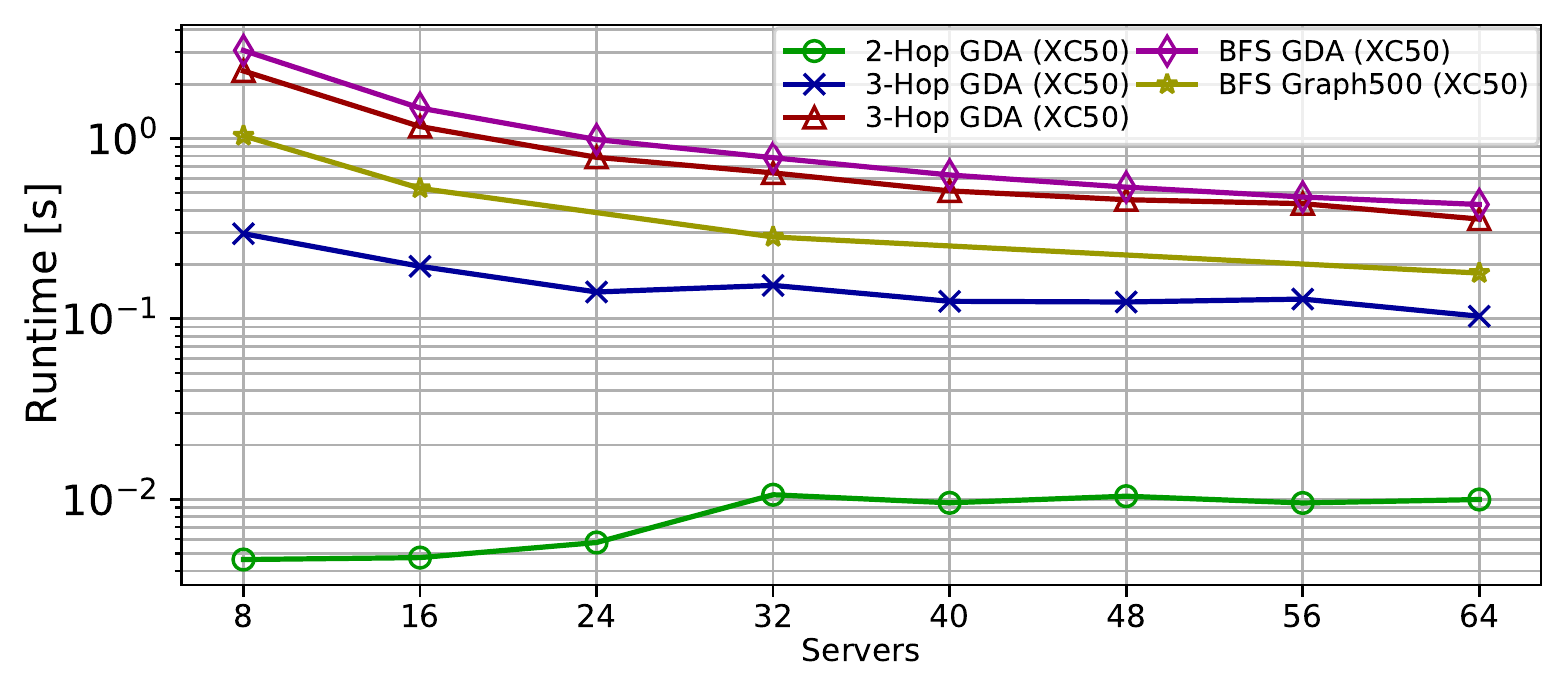}
\vspaceSQ{-1.5em}
\caption{\textmd{BFS, k-hop; strong scaling.}}
\label{fig:olap-bfs-khop-strong-scaling}
\end{subfigure}
\vspaceSQ{-0.5em}
\caption{\textmd{Analysis of OLAP and OLSP workloads. \textbf{PR}: PageRank,
\textbf{CDLP}: Community Detection using Label Propagation, \textbf{WCC}:
Weakly Connected Components, \textbf{LCC}: Local Cluster Coefficient,
\textbf{BI2}: Business Intelligence 2 query from LDBC SNB,
\textbf{GNN}: Graph Neural Networks (training of the graph convolution model), \textbf{weak scaling}: scaling
dataset sizes together with \#servers, \textbf{strong scaling}: scaling
\#servers for a fixed dataset (a Kronecker graph of scale 26, i.e., 67.1M
vertices and 1.1B edges; the results follow the same performance patterns for
other datasets).
\textbf{Missing data points} of our baselines indicate limited compute budget;
missing baselines of comparison targets indicate inability to scale to a given
configuration; isolated GDA data points not connected with lines to the rest of
the data series indicate extreme-scale runs.}}
\label{fig:eval-olap}
\vspaceSQ{-1em}
\end{figure*}

\subsection{Selecting Baselines and Related Challenges}

While there exist many graph databases, the vast majority of them is
not freely available.
We attempted to get access to different systems, such as Oracle's PGX,
but our attempts were unsuccessful.
%
%
%
Among the available systems, we shortlisted databases that provide full support
for both OLTP and OLAP queries.
After an extensive investigation and configuration effort, we were able to
successfully configure and use Neo4j (5.10)~\cite{neo4j_book} and
JanusGraph (0.6.2)~\cite{janus_graph_links}.
We configure both baselines for in-memory execution. Additionally, to
maximize the performance of comparison baselines, we use their
high-performance consistency guarantees (e.g., eventual consistency for
JanusGraph), even if GDI provides serializability for graph updates.
\emph{These two systems are two of the highest-ranking core graph databases (i.e.,
systems with the database model ``Graph'') in the DB-Engines Ranking.}

\subsection{Distributed In-Memory LPG Graph Generator for Massive-Scale Experiments}

Obtaining appropriate {graph datasets} is challenging due to the fact that we
target graphs of very large scales \emph{and} having rich amounts of labels and
properties. 
\iftr
Even the largest publicly available graph, Web Data
Commons~\cite{wdc}, only has 128B edges and no labels/properties.
Moreover, the LDBC data generator experienced regular OOM problems, whenever we
tried to generate graphs of very large scales.
Another problem with both real-world graphs and with the graphs
generated by the LDBC generator is that they are stored in the distributed
filesystem storage. Loading such large graph from disks is time-consuming and
stresses compute budget. 
\fi
Hence, to facilitate large-scale graph database experiments, we develop an
\emph{in-memory distributed generator of LPG graphs} that enables \emph{fast}
construction of \emph{arbitrarily large LPG datasets limited only by the
available compute resources}, fully in-memory, so that they are immediately
available for further processing. We base our generator on the existing code
provided by the Graph500 benchmark~\cite{graph500} that uses the realistic
Kronecker random graph model with a heavy-tail skewed degree
distribution~\cite{leskovec2010kronecker}.
\iftr
In this model, one specifies two
input parameters, the graph vertex scale $s$ and the edge factor $e$.  The
resulting graph has $2^s$ vertices and approximately $e$ edges per vertex, for
the total of $e 2^s$ edges.
\fi
We extend this model by adding support for a user-specified selection (i.e.,
counts and sizes) of labels and properties, and how they are assigned to
vertices and edges.
%
%
By default, we use 20 different labels and 13 property types in the
following analyses (we also experiment with varying these values).
\iftr
Internally, the vertices are distributed round robin, which is an
implementation detail that can be modified by the GDA developer.
\fi

\subsection{Analysis of OLTP Workloads}

We first analyze the OLTP workloads. Here, we stress GDA with a high-velocity
stream of graph queries and transactions.
We use four specific scenarios based on the {LinkBench
benchmark}~\cite{armstrong2013linkbench} and on other past GDB
evaluations~\cite{dubey2016weaver,chen2021g}, see Table~\ref{tab:oltp-workloads} for
details.

\begin{table}[b]
\vspaceSQ{-1.5em}
\centering
\setlength{\tabcolsep}{2pt}
\renewcommand{\arraystretch}{0.7}
\footnotesize
\scriptsize
\begin{tabular}{lllll@{}}
\toprule
\textbf{Operation} & 
\makecell[l]{\textbf{``Read Mostly''}\\ \textbf{(RM)}~\cite{dubey2016weaver}} & 
\makecell[l]{\textbf{``Read Intensive''}\\ \textbf{(RI)}~\cite{dubey2016weaver}} & 
\makecell[l]{\textbf{``Write Intensive''}\\ \textbf{(WI)}~\cite{chen2021g}} &
\makecell[l]{\textbf{LinkBench}\\ \textbf{(LB)}~\cite{armstrong2013linkbench}} \\ 
\midrule
\textbf{Read queries:} & \textbf{99.8\%} & \textbf{75\%} & \textbf{20\%} & \textbf{69\%} \\
Get vertex properties & 28.8\% & 21.7\% & 9.1\% & 12.9\% \\ 
Count edges of a vertex & 11.7\% & 8.8\% & 0\% & 4.9\% \\
Get edges of a vertex & 59.3\% & 44.5\% & 10.9\% & 51.2\% \\
\midrule
\textbf{Update queries:} & \textbf{0.2\%} & \textbf{25\%} & \textbf{80\%} & \textbf{31\%} \\
Add a new vertex & 0\% & 0\% & 20\% & 2.6\% \\
Delete a vertex & 0\% & 0\% & 6.7\% & 1\% \\
Update a vertex property & 0\% & 0\% & 13.3\% & 7.4\% \\
Add a new edge & 0.2\% & 25\% & 40\% & 20\% \\
\bottomrule
\end{tabular}
\caption{OLTP workloads described in this paper. We varied the fractions of
specific types of operations for broad investigation beyond the ones provided
here, all results followed similar patterns to those described here.}
\vspaceSQ{-3em}
\label{tab:oltp-workloads}
\end{table}

We first evaluate the overall throughput, see Figure~\ref{fig:eval-oltp}.  GDA
achieves high scalability: adding more servers consistently improves the
throughput in both strong and weak scaling. 
%
%
Throughput increase is particularly visible in the RI and RM workloads with
more read queries, because LB and WI workloads come with more updates that
involve more synchronization and communication.
We also observe that, overall, XC50 servers give more performance than XC40,
especially for RM workloads dominated by reads. We conjecture this is due to
the XC50 servers offering more network bandwidth per core.
Moreover, we note that very low percentages of failed transactions (less than
0.2\% for RI/RW and less than 2\% for LB/WI) across all benchmarks indicate
GDA's capability to successfully resolve a sustained stream of incoming user
requests, even at very high scales.
Overall, the results indicate that GDA is able to both accelerate requests into
a given fixed dataset (as seen by the throughput increase in strong scaling) as
well as it enables scaling to larger datasets (as indicated by the throughput
increase in weak scaling).

We also show histograms of latencies of different operations within a given
OLTP workload. Figure~\ref{fig:eval-oltp-dets} shows the data for LB
(we plot separate latencies for transactions running on 1--8 servers).
GDA is consistently the fastest, with the vast
majority of its operations being below 1$\mu$s (for 1 server) and close to 10--100$\mu$s
(for more servers), even for demanding vertex deletions. 
JanusGraph requires at least 500$\mu$s for all the operations (in most 
of cases), with no operation being faster than 200$\mu$s, even for the single
server scenario. Vertex deletions start at around 2000$\mu$s.
Our advantages are even more distinctive considering the fact that GDA ensures
serializability, while JanusGraph uses its default configuration with a more
relaxed eventual consistency. 
Neo4j is slower than both GDA and JanusGraph; it however shows similar
trends in the differences between particular operation types (e.g., read
operations are on average faster than updates). While most Neo4j operations
finish below 20ms, it does entail relatively many outliers.

\textbf{Summary of GDA's Advantages}
GDA is faster than comparison targets due to its fundamental reliance
on one-sided RDMA.

\enlargeSQ

\subsection{Analysis of OLAP and OLSP Workloads}

We illustrate the OLAP and OLSP results in Figure~\ref{fig:eval-olap}.
We consider BFS, PageRank (PR), Community Detection using Label Propagation
(CDLP), Weakly Connected Components (WCC), Local Cluster Coefficient (LCC),
Business Intelligence~2 query from LDBC SNB
(BI2)~\cite{DBLP:journals/pvldb/SzarnyasWSSBWZB22}, and Graph Neural Networks
(GNN; training of the graph convolution model~\cite{kipf2016semi}).
The results follow advantageous performance patterns -- for most problems (BFS,
k-hop, GNN) adding more compute resources combined with increasing the dataset
size only results in mild runtime increases (in weak scaling) or runtime drops
(in strong scaling). WCC, CDLP, and PR are characterized by overall sharper slopes
of increasing running times for weak scaling; we conjecture this is because
these problems cumulatively involve more communication due to their memory
access patterns and runtime complexities (e.g., LCC has the complexity of $O(n
+ m^{3/2})$ compared to $O(m+n)$ for BFS).

\enlargeSQ

\iftr
We compare GDA to a very competitive target, the Graph500 implementation of
BFS~\cite{graph500}. It is a highly tuned BFS code
\fi
that has been used for many years to assess high-performance clusters in their
abilities to process graph traversals. Graph500 uses graphs with no labels
or properties, and it does not use graph transactions.
Importantly, GDA is at most 2--4$\times$ slower than Graph500, and 
sometimes it is comparable or even faster (e.g., see 2,048 servers for weak
scaling). Hence, GDA is able to deliver high performance
graph analytics of even largest scales considered.

\textbf{Summary of GDA's Advantages}
Using MPI collectives gives GDA significant benefits for OLAP/OLSP queries.
As collectives offer clear semantics, they further boost performance by
eliminating boilerplate code.

\enlargeSQ

\subsection{Varying Labels, Properties, \& Edge Factors}

In addition to scaling graph sizes (\#vertices and \#edges), we also analyze
GDA's performance for graphs with different amounts of labels and properties.
Intuitively, graphs with very few of these have little rich data attached to
vertices and edges. Thus, workloads are mostly dominated by irregular
distributed memory single-block reads and writes. With more labels and
properties, data accesses are still irregular (due to the nature of graph
workloads), but reads and writes may access many blocks.
GDA's advantages are preserved in all these cases, thanks to harnessing the
underlying RDMA.

We use the default value of the edge factor $e=16$, which results in Kronecker
graphs close to many real-world datasets in terms of their degree distribution
and sparsity. We also tried other values of $e$, they also come with similar
GDA's advantages.

\subsection{Analysis of Real-World Graphs}

We also consider large real-world graphs, 
(which includes Web Data Commons and other largest publicly available
real-world datasets) selected from the
KONECT~\cite{kunegis2013konect}
and WebGraph~\cite{boldi2004webgraph} repositories.
The performance patterns and GDA's advantages are similar to those obtained for
Kronecker graphs.
This is because both the considered real-world and Kronecker graphs have
similar sparsities as well as heavy-tail degree distributions that have been
identified as key factors that determine performance patterns.
For example, we were able to process an OLAP BFS query on the Web Data Commons
dataset with $\approx$3.56 billion vertices and $\approx$128 billion edges in
$\approx$15s using 1,024 XC50 servers.

\enlargeSQ

\subsection{Extreme Scales \& Comparison to Others}

Our evaluation comes with the largest experiments described in the literature
in terms of \#servers, \#cores, and \#edges.
These largest runs are pictured in Figure~\ref{fig:oltp-read-weak-scaling} for
OLTP (RM), and in Figures~\ref{fig:olap-global-weak-scaling}
and~\ref{fig:olap-bfs-khop-weak-scaling} for OLAP (WCC, PR, BFS).  We were only
able to run a few such experiments due to the fact that it required using the
full scale of the Piz Daint supercomputer.
The results illustrate that even at such workloads, GDA still offers high
scalability. For example, moving from a graph with 275B edges to 550B edges
increases the OLTP throughput by $\approx$3$\times$ while
\#servers increase by 3.49$\times$.

One recent study with large-scale executions is from the commercial ByteGraph
system~\cite{libytegraph}.  However, it does not specify the details of the
used graph, and it partially uses disks.
Second, while a recent study of the TigerGraph commercial system comes with a
graph of a similar size to us (539.6B edges)~\cite{tigergraph2022ldbc}, their servers have significantly
more memory (each has $\approx$1TB vs.~64 GB in our setting).  As the network
is the main bottleneck in large-scale communication, we expect that GDA would
also scale well with such fat-memory servers, and thus it could be able to
scalably process even larger graphs than the ones we tried.

Finally, note that our runs required using both XC40 and XC50 servers
simultaneously to use full Piz Daint's scale. As XC40 and XC50 come with
different CPUs and core counts, this may cause load imbalance. Thus, we
conjecture that when using GDA in production data centers with uniform servers,
its performance and scalability could be even better than described in this
work.

\iftr\section{PORTABILITY DISCUSSION}
\label{sec:port}

Our work fosters two types of portability.
%
%
First, we incorporate a similar type of portability as MPI. Namely, if a given
graph database implementation adheres to the GDI specification, and if there is
an available implementation of GDI for a given architecture, then the
implementation would seamlessly compile and execute.
Further, GDI is fully independent of the architecture details and it thus can
be seamlessly implemented on x86, ARM, and others.
%
%
While facilitating GDI's adoption, this would still require separate GDI implementations
for different RDMA architectures, e.g., Cray, IBM, or InfiniBand.
To alleviate this, we use plain RMA operations (puts, gets, atomics) to implement GDI-RMA.
Such operations are broadly supported by RDMA networks.
This facilitates porting the GDI-RMA code to other RDMA settings.

\fi
\section{RELATED WORK}

\iftr
\textbf{GDBs} 
have been researched in both academia and
industry~\cite{angles2018introduction, davoudian2018survey, han2011survey,
gajendran2012survey, gdb_survey_paper_Kaliyar, kumar2015domain,
gdb_survey_paper_Angles}, in terms of query languages~\cite{angles2018g,
bonifati2018querying, gdb_query_language_Angles, ten2023duckpgq}, database
management~\cite{gdb_management_huge_unstr_data, pokorny2015graph,
junghanns2017management, bonifati2018querying, miller2013graph}, compression
and data models~\cite{lyu2016scalable, ma2016big, nabti2017querying,
besta2018survey, besta2019slim, besta2018log, besta2022probgraph}, execution in
novel environments such as the serverless setting~\cite{toader2019graphless,
copik2020sebs, mao2022ermer}, and others~\cite{deutsch2020aggregation}.  Many
graph databases exist~\cite{tigergraph2022ldbc, cge_paper, tiger_graph_links,
janus_graph_links, chen2019grasper, azure_cosmosdb_links, amazon_neptune_links,
virtuoso_links, arangodb_links, tesoriero2013getting, profium_sense_links,
triplebit_links, gbase_paper, graphbase_links, graphflow, livegraph,
memgraph_links, dubey2016weaver, sparksee_paper, graphdb_links,
redisgraph_links, dgraph_links, allegro_graph_links, apache_jena_tbd_links,
mormotta_links, brightstardb_links, gstore, anzo_graph_links, datastax_links,
infinite_graph_links, blaze_graph_links, oracle_spatial, stardog_links,
cayley_links}.
In this context, GDI offers standardized building blocks for GDBs to
foster portability and programmability across different architectures.
\fi

Many \textbf{workload specifications and benchmarks for GDBs} exist,
covering OLTP interactive queries (SNB~\cite{ldbc_snb_specification},
LinkBench~\cite{armstrong2013linkbench}, and BG~\cite{barahmand2013bg}), OLAP
workloads (Graphalytics~\cite{ldbc_graphanalytics_paper}), or business
intelligence queries (BI~\cite{DBLP:journals/pvldb/SzarnyasWSSBWZB22,
early_ldbc_paper}).
One can express these workloads using portable and programmable GDI building
blocks.
\iftr
Moreover, there are many evaluations of GDBs~\cite{capotua2015graphalytics,
dominguez2010survey, mccoll2014performance, jouili2013empirical,
ciglan2012benchmarking, lissandrini2017evaluation, lissandrini2018beyond,
tian2019synergistic}. We complement these works by providing a large-scale
RDMA-focused evaluation.
\fi
Note that global analytics workloads are the focus of Pregel-like
systems~\cite{malewicz2010pregel, gonzalez2012powergraph, chen2019powerlyra}.
These systems are mostly incomparable to GDBs because they do not support graph
updates or LPG datasets.

\iftr

\textbf{RDMA \& RMA}
RDMA has been one of the enablers of high performance on supercomputers and
data centers, and is widely supported~\cite{karmarkar2015availability,
alibaba_cloud, oracle_cloud, roce, iwarp, arimilli2010percs, IBAspec}.
Numerous libraries and languages offer RMA features. Examples include MPI-3
RMA~\cite{mpi3}, UPC~\cite{upc}, Titanium~\cite{hilfinger2005titanium}, Fortran
2008~\cite{fortran2008}, X10~\cite{x10}, or Chapel~\cite{chapel}. 
We target RDMA and RMA systems with our implementation.

\fi

\textbf{Resource Description Framework (RDF)}~\cite{lassila1998resource} is a
standard to encode knowledge in ontological
models and in RDF stores using triples~\cite{modoni2014survey, harris20094store,
papailiou2012h2rdf}.
\iftr
There exists a lot of works on RDF and knowledge
graphs~\cite{ristoski2016rdf2vec, portisch2020rdf2vec, jurisch2018rdf2vec,
ristoski2019rdf2vec, lin2015learning, arora2020survey, wang2017knowledge,
dai2020survey, urbani2016kognac, shi2017proje, trouillon2017knowledge,
akrami2020realistic, yao2019kg, sun2019re, chen2020knowledge, lin2015learning,
shi2018open}.
\fi
%
%
We focus on graph databases built on top of LPG, and thus RDF designs are
outside the scope of this work.

\iftr

\textbf{Dynamic, Temporal, and Streaming Graph Frameworks}
There are many systems that process dynamic, temporal, and streaming
graphs~\cite{besta2019practice, sakr2021future, choudhury2017nous}.
Their design overlaps with graph databases, because they also focus on
high-performance graph queries and updates, and on solving global analytics.
Examples include Betweenness Centrality~\cite{pontecorvi2015faster, ediger2012stinger,
tsalouchidou2020temporal, solomonik2017scaling, madduri2009faster}, Graph
Traversals~\cite{ediger2012stinger, macko2015llama, sengupta2017evograph,
sengupta2016graphin, besta2015accelerating, besta2017slimsell,
kepner2016mathematical, bulucc2011combinatorial}, Connected
Components~\cite{macko2015llama, sengupta2016graphin, ediger2012stinger,
feng2020risgraph, feng2020risgraph, gianinazzi2018communication}, Graph
Coloring~\cite{duan2019dynamic, besta2020highcolor},
Matchings~\cite{bernstein2021deamortization, neiman2015simple,
besta2020substream}, and many others~\cite{thorup2000near, lee2016efficient,
henzinger1999randomized, demetrescu2009dynamic, eppstein1999dynamic,
gianinazzi2021parallel, ivkovic1993fully, besta2019communication,
besta2017push}.
Recent efforts work towards processing
subgraphs~\cite{qin2019mining} such as maximal cliques or dense
clusters~\cite{cook2006mining, jiang2013survey, horvath2004cyclic,
chakrabarti2006graph, besta2021sisa, gms, besta2022probgraph}.
However, such frameworks usually do not focus on rich data, do not use the full LPG
model, and rarely consider ACID properties, thus being outside the focus of
this paper.

\iftr
\textbf{Graph Processing APIs}
There are efforts for
developing APIs different parts of the graph processing landscape.
Importantly, GraphBLAS
is a standard for formulating graph algorithms using linear algebra
building blocks~\cite{kepner2016mathematical, kepner2015graphs}.
It
targets shared-memory CPU environments
(GraphMat~\cite{sundaram2015graphmat, mattson2014standards}), shared-memory GPU
settings (GraphBLAST~\cite{yang2022graphblast}), and distributed-memory
environments (CombBLAS~\cite{azad2021combinatorial} and others~\cite{DBLP:conf/ipps/BrockBMMMPSS20}).
The difference is that these APIs focus on analytics on basic graph models (simple
graph/multi-graph/weighted graphs) and on incorporating sparse
linear algebra as the main building blocks.
GraphBLAS routines could be used together with GDI. For example, the OLAP analytics
supported by GDI could be implemented using GraphBLAS codes.
\fi

\textbf{Graph Learning, Graph Neural Networks, Neural Graph Databases}
Graph neural networks (GNNs) have recently become an established part of the
deep learning landscape~\cite{wu2020comprehensive, zhou2020graph,
zhang2020deep, chami2020machine, hamilton2017representation,
bronstein2017geometric, hamilton2020graph, besta2021motif,
gianinazzi2021learning, scarselli2008graph, besta2022parallel,
mirhoseini2021graph, jumper2021highly, pfaff2020learning, sanchez2020learning,
davies2021advancing}.
The versatility of GNNs brings a promise of enhanced analytics capabilities
in the graph database landscape.
Recently, {Neo4j Inc., Amazon, and others} have started to investigate how to
use GNNs within their systems~\cite{hodlergraph}. Here, Neo4j already supports
obtaining embeddings and using them for node or graph classification. However,
there is still limited support for taking advantage of the full scope of
information provided by LPG labels/properties.
This has been alleviated by LPG2vec~\cite{besta2022neural}, an architecture for
enabling Neural Graph Databases based on LPG.
While in this work we do not focus specifically on GNNs, we illustrate that GDI
can also be used to support graph ML workloads in the GDB domain, and thus it
can serve as a propeller for the upcoming Neural Graph Database systems.

\fi

\section{CONCLUSION} 


Graph databases (GDBs) are of central importance in academia and industry, and
they drive innovation in many domains ranging from computational chemistry to
engineering. However, with extreme-scale graphs on the horizon, they face
several challenges, including high performance, scalability, programmability,
and portability. 

In this work, we provide the first systematic approach to address these
challenges. First, we design the Graph Database Interface (GDI): an
MPI-inspired specification of performance-critical building blocks for the
transactional and storage layer of a GDB. 
By incorporating the established MPI principles into the GDB domain, we enable
designing GDBs that are portable, have well-defined behavior, and seamlessly
incorporate workloads as diverse as OLTP, OLAP, and OLSP.
Moreover, while in the current GDI release we focus on Labeled Property Graphs,
GDI can be straightforwardly extended to cover other data models such as RDF or
Knowledge Graphs.
This would further illustrate the applicability of HPC-based design principles
even well beyond traditional graph databases.

\enlargeSQ

To illustrate the potential of GDI in practice, we use it to build GDI-RMA, a
graph database for distributed-memory RDMA architectures. 
To the best of our knowledge, GDI-RMA is the first GDB that
harnesses many powerful HPC mechanisms, including collective communication,
offloaded RDMA, non-blocking communication, and network-accelerated
atomics.
We crystallize the most important design decisions into generic
comprehensive insights that can be reused for developing other high-performance
GDBs.

In evaluation, we achieve unprecedented performance and scalability for a
plethora of workloads, including diverse small transactional
queries as well as large graph analytics such as Community Detection or Graph
Neural Networks. GDI-RMA outperforms not only other graph databases by orders
of magnitude, but its implementation of BFS approaches and even matches in some
cases Graph500, a high-performance graph traversal implementation tuned over
many years.  This is an important result, because the Graph500 kernel only
implements the single BFS algorithm for static simple graphs without any rich
data, while GDI-RMA is a GDB engine with transactional support for arbitrary
graph modifications, the LPG rich data model, and many types of queries.

Finally, we deliver the largest experiments reported in the GDB literature in
terms of \#servers and \#cores, improving upon previous results by orders of
magnitude.
Our code is publicly available and can be used to propel developing
next-generation GDBs.

\vspace{1em}


%

\iftr

{
\section*{ACKNOWLEDGEMENTS}
%
We thank Hussein Harake, Colin McMurtrie, Mark Klein, Angelo Mangili, Marco
Induni, Andreas Jocksch, Maria Grazia Giuffreda and the
whole CSCS team granting access to the Ault and Daint machines and for their
excellent technical support.
We thank Timo Schneider for immense help with infrastructure at SPCL,
and PRODYNA AG (Darko Križić, Jens Nixdorf, Christoph Körner) for generous
support.
We thank Emanuel Peter, Claude Barthels, Jakub Ja\l{}owiec,
Roman Haag, Jan Kleine, Lasse Meinen, and Janez Ales (BASF SE) for their help with the project.
This project received funding from the European Research Council
\raisebox{-0.25em}{\includegraphics[height=1em]{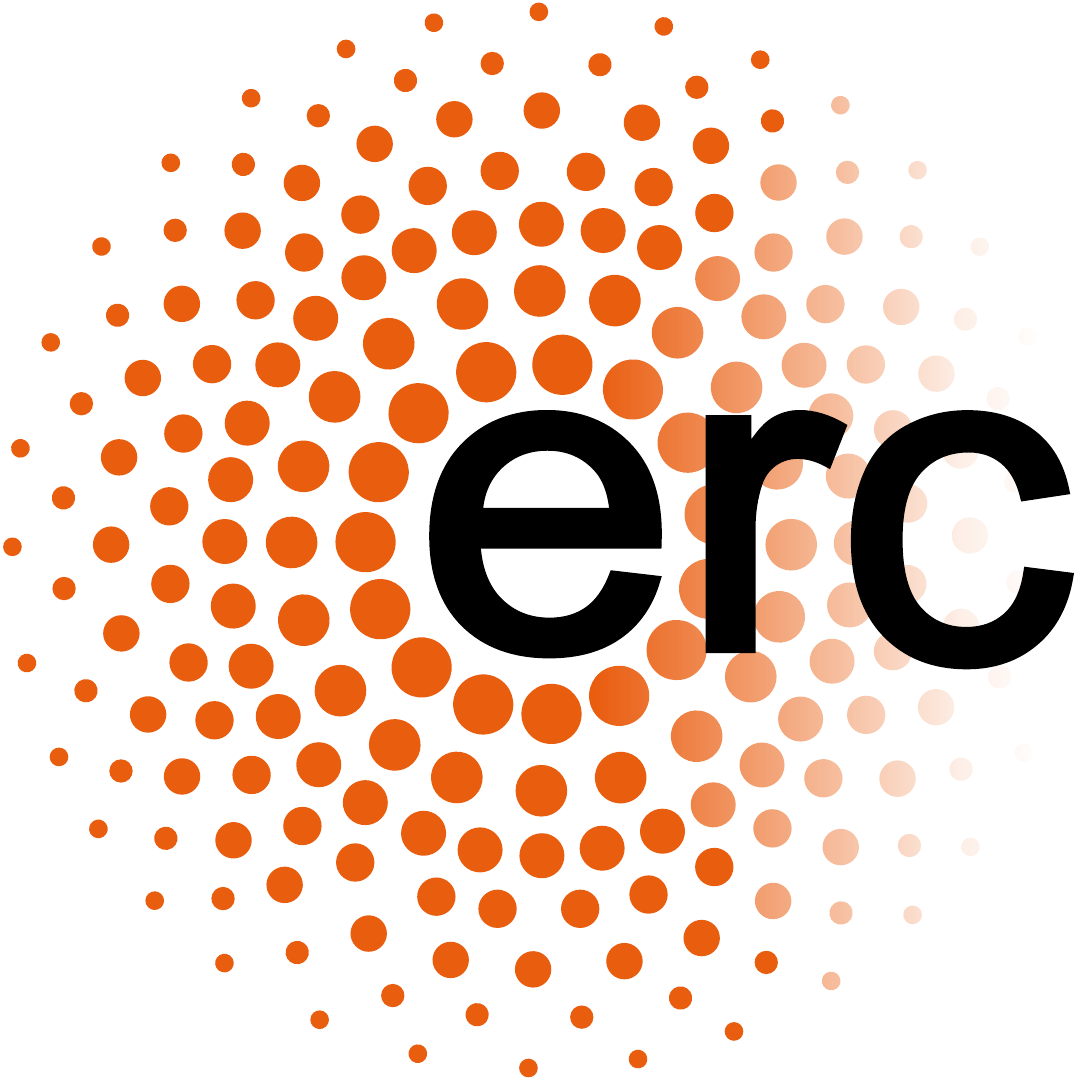}} (Project PSAP,
No.~101002047), and the European High-Performance Computing Joint Undertaking
(JU) under grant agreement No.~955513 (MAELSTROM).
This project was supported by the ETH Future Computing Laboratory (EFCL),
financed by a donation from Huawei Technologies.
This project received funding from the European Union's HE research and
innovation programme under the grant agreement No.~101070141 (Project
GLACIATION).
This work was also supported in part by the European Union's Horizon 2020
research and innovation programme under the grant: Sano No.~857533 and the
International Research Agendas programme of the Foundation for Polish Science,
co-financed by the EU under the European Regional Development Fund.
}

\fi

\bibliographystyle{ACM-Reference-Format}
\bibliography{references.complete}

\end{document}